\documentclass[aps,prb,showpacs,twocolumn,floats,epsfig,pdflatex, superscriptaddress]{revtex4-2}
\usepackage[T1]{fontenc}
\usepackage[latin9]{inputenc}
\usepackage{color}
\definecolor{shadecolor}{rgb}{1, 0, 0}
\usepackage{verbatim}
\usepackage{float}
\usepackage{framed}
\usepackage{amsmath}
\usepackage{amssymb}
\usepackage{gensymb}
\usepackage{graphicx}
\usepackage{physics}
\usepackage{siunitx}
\usepackage[normalem]{ulem}

\usepackage{slashed}
\usepackage{subfigure}

\makeatletter
\PassOptionsToPackage{caption=false}{subfig}
\usepackage{hyperref}
\hypersetup{
breaklinks=true,
colorlinks=true,
citecolor=blue,
linkcolor=blue,
filecolor=blue,
urlcolor=blue
}
\IfFileExists{lmodern.sty}{\usepackage{lmodern}}{}

\makeatother

\usepackage{babel}

\begin{document}

\title{Optimizing superlattice bilayer graphene for a fractional Chern insulator}

\author{Dathan Ault-McCoy}
\affiliation{Department of Physics and Astronomy, Stony Brook University, Stony Brook, New York 11794, USA}

\author{M. Nabil Y. Lhachemi}
\affiliation{Department of Physics and Astronomy, Stony Brook University, Stony Brook, New York 11794, USA}

\author{Aaron Dunbrack}
\affiliation{Department of Physics and Nanoscience Center, University of Jyv\"askyl\"a, P.O. Box 35, FI-40014 University of Jyv\"askyl\"a, Finland}
\affiliation{Department of Physics and Astronomy, Stony Brook University, Stony Brook, New York 11794, USA}

\author{Sayed Ali Akbar Ghorashi}
\affiliation{Department of Physics and Astronomy, Stony Brook University, Stony Brook, New York 11794, USA}

\author{Jennifer Cano}
\affiliation{Department of Physics and Astronomy, Stony Brook University, Stony Brook, New York 11794, USA}
\affiliation{Center for Computational Quantum Physics, Flatiron Institute, New York, New York 10010, USA}

\date{\today}                                       

\begin{abstract}
Bernal-stacked bilayer graphene modulated by a superlattice potential is a highly tunable system predicted to realize isolated topological flat bands. In this work we calculate the band structure and quantum geometry of bilayer graphene subject to both triangular and square superlattices, across a wide range of gate voltages. We identify the parameter regime that optimizes the ``single-particle indicators'' for the stability of a fractional Chern insulator (FCI) when a topological flat band is partially filled. Our results guide the experimental realization of an FCI in this platform.
\end{abstract}
\maketitle

\section{Introduction}

The discovery of superconductivity and correlated insulators in twisted bilayer graphene (TBG) launched the field of moir\'e materials \cite{bistritzer2011,Morell2010,cao2018correlated,cao2018unconventional}.
Moir\'e materials generically exhibit flat bands as a consequence of an emergent, nanometer-scale moir\'e superlattice.
Often, these flat bands inherit topology from their constituent layers \cite{tarnopolsky2019origin,wu2019topological,pan2020band,devakul2021magic,pan2022topological,zhang2021spin,crepel2025efficient}, which is responsible for the observation of the anomalous Hall effect in both TBG \cite{serlin2020intrinsic,sharpe2019emergent} and twisted transition metal dichalcogenide (TMD) heterobilayers \cite{li2021quantum}.
In a finite magnetic field, Chern insulators -- lattice-analogs of quantum Hall states -- have also been observed
in TBG \cite{nuckolls2020strongly,xie2021fractional}, building on earlier observations in graphene heterostructures aligned with hexagonal boron nitride (hBN) \cite{dean2013moire,hunt2013massive,ponomarenko2013placing,spanton2018}. 

Despite these observations, the long sought-after zero-field fractional Chern insulator (FCI) \cite{regnault2011fractional,sheng2011fractional,neupert2011fractional} has remained elusive in TBG.
Instead, the first observation of a zero-field FCI appeared recently in twisted MoTe\textsubscript{2} \cite{cai2023,park2023,zeng2023}.
The second observation followed shortly after in rhombohedral graphene aligned with hBN \cite{lu2024fractional,xie2025tunable}.
These two observations in diverse experimental platforms give optimism that more realizations will soon follow.

\begin{figure}
    \centering
    \includegraphics[width=\linewidth]{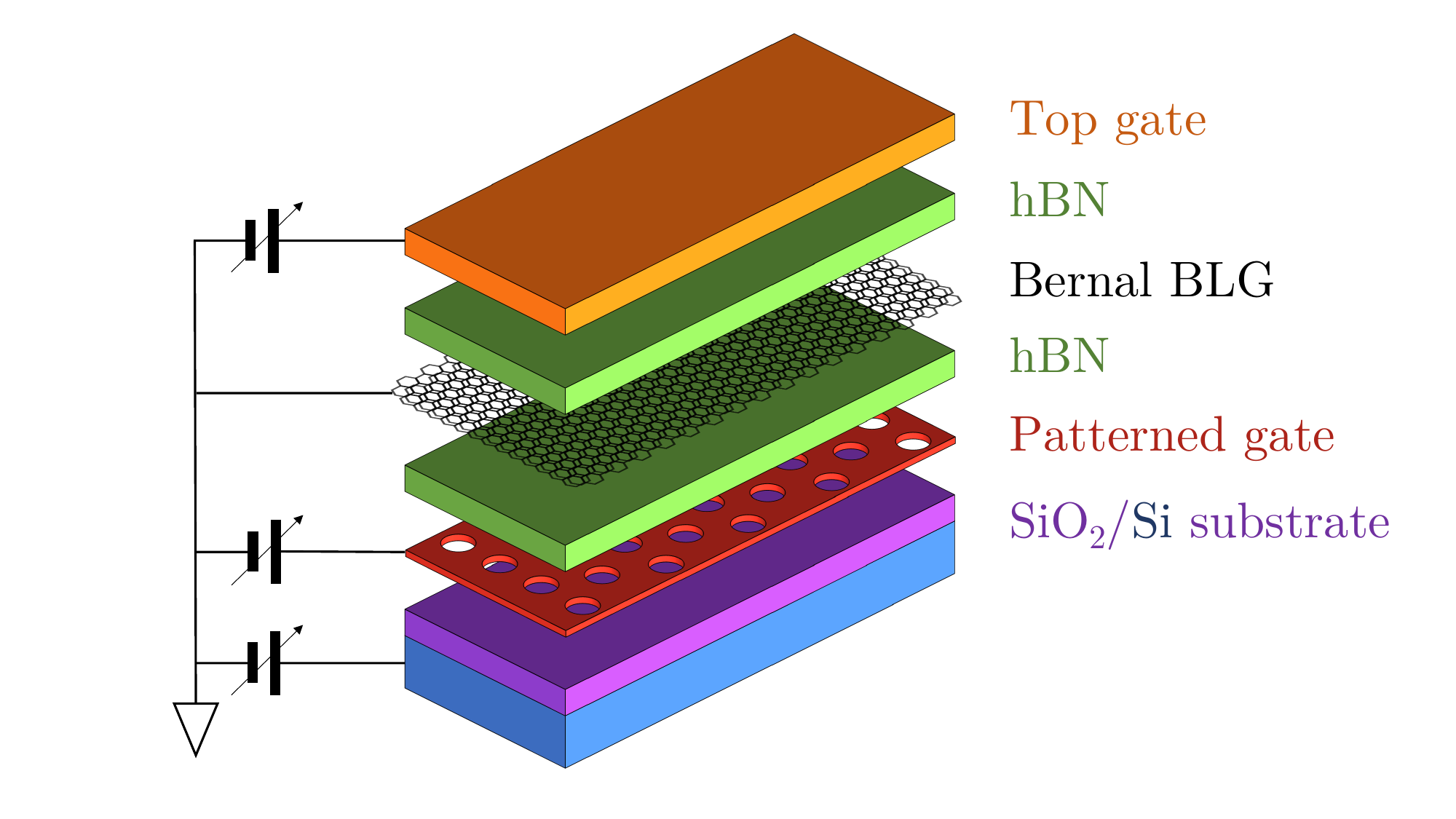}
    \caption{Schematic of an experimental set-up realizing sBLG.
    A spatially modulated electric potential is generated on Bernal-stacked bilayer graphene by a nanopatterned bottom gate. Along with uniform top and bottom gates, the spatially modulated potential, spatially constant displacement field and the electron density can all be independently tuned.
    }
    \label{fig:exp-setup}
\end{figure}

Recently, experimental advances in gate-patterning \cite{forsythe2018band, wang2018observation, li2021anisotropic, barcons2022engineering,wang2023formation,sun2024signature,wang2024tuning} and moir\'e \cite{yasuda2021stacking, vizner2021interfacial, wang2022interfacial,kim2024electrostatic,zhang2024engineering,wang2025moire} and molecular engineering \cite{gomes2012designer} have inspired an increasingly diverse array of proposals for new systems with designer superlattice geometry beyond the twisted bilayers \cite{shi2019gate,superlattice,krix2023patterned,crepel2023chiral,gao2023untwisting,wan2023topological,zeng2024gate}.
In this paper, we study one such system which has been put forward as a promising platform to realize an FCI \cite{superlattice,zeng2024gate}: Bernal-stacked bilayer graphene subject to a spatially periodic electric potential, henceforth referred to as superlattice bilayer graphene (sBLG).
This system can be realized by the experimental setup depicted in Fig.~\ref{fig:exp-setup}, where the superlattice potential is defined by a patterned electrode. This platform may avoid some practical challenges of moir\'e heterostructures \cite{lau2022reproducibility}, while offering independent control over the superlattice geometry, length scale, and potential strength, the last of which is tunable \textit{in situ}. 

While experimental evidence for correlated phenomena in sBLG has been observed and attributed to superlattice-induced flat bands \cite{sun2024signature}, the theoretically predicted topological regime has not been realized.
The purpose of the present work is to systematically characterize the parameter space of sBLG and determine the optimal parameters to realize an FCI. As a proxy for FCI stability, we employ single-particle indicators based on quantum band geometry, which we will describe shortly. Our work serves as a guide for future experimental studies of this platform.

\section{Model}
\label{sec:model}

We model the band structure of sBLG by the Hamiltonian
\begin{equation}
    H = H_{\text{BLG}} + H_V.
\end{equation}
The first term describes an effective continuum model for bilayer graphene derived by expanding a tight binding model around the $K$ or $K'$ high-symmetry point 
\cite{mccann2013electronic}, and is given by 
$H_{\text{BLG}} = \sum_{\vb{k}} c^\dagger_{\vb{k}} \mathcal{H}_{\text{BLG}}(k) c_{\vb{k}}$, 
where $c_{\vb{k}} = \begin{pmatrix} c_{\vb{k},A1} & c_{\vb{k},B1} & c_{\vb{k},A2} & c_{\vb{k},B2} \end{pmatrix}^T$,
the operator $c_{\vb{k},Ij}$ annihilates an electron with quasimomentum $\vb{k}$ on sublattice $I$ and layer $j$ (with $j = 1$ corresponding to the top layer in Fig.~\ref{fig:exp-setup}),
and
\begin{equation} \label{eq:Hblg} 
   \mathcal{H}_{\text{BLG}}(k) = \begin{pmatrix} 
        0 & vk^* & -v_4 k^* & v_3 k \\
        vk & 0 & t & -v_4 k^* \\
        -v_4 k & t & 0 & vk^* \\
        v_3 k^* & -v_4 k & v k & 0
    \end{pmatrix}. 
\end{equation}
Here, $k = k_x + i \chi k_y$ is the complexified quasimomentum with $\chi \in \{+,-\}$ the valley index.
In the remainder of this work, we will specialize to the $K$ valley, taking $\chi=+$.
In Eq.~(\ref{eq:Hblg}), $v$ is the Fermi velocity, $t$ is the nearest-neighbor inter-layer hopping strength, and $v_3$ and $v_4$ arise from next-to-nearest-neighbor hopping contributions. These latter two terms result in trigonal warping of the Fermi surface and electron-hole asymmetry, respectively.
Though they were omitted in previous work, e.g., Ref.~\cite{superlattice}, for the purposes of the present work, we find their effects quantitatively significant enough to warrant inclusion.

The electric potential applied to the sample is included through
\begin{equation}
    H_V = \int \dd{\vb{r}} \sum_{I,j} \psi_{Ij}^\dagger(\vb{r}) \psi_{Ij}(\vb{r}) \qty[(-1)^j V_0 + \alpha_j V_{\text{SL}}(\vb{r})],
\end{equation}
where $\psi_{Ij}(\vb{r}) = \sum_{\vb{k}} c_{\vb{k},Ij} e^{i\vb{k} \cdot \vb{r}}$, $V_0$ is a uniform interlayer displacement potential, and $V_{\text{SL}}(\vb{r})$ is the spatially modulated part of the potential. The symbol $\alpha_j$ is introduced to capture the screening effect of the second graphene layer on the potential felt by the first, and is defined by $\alpha_1 = \alpha$ and $\alpha_2 = 1$, where $\alpha \leq 1$ is a parameter of the model. We further assume that $V_{\text{SL}}(\vb{r})$ is simple harmonic and so can be written as
\begin{equation}
    V_{\text{SL}}(\vb{r}) = V_{\text{SL}} \sum_n \cos(\vb{Q}_n \cdot \vb{r}),
    \label{eq:VSL}
\end{equation}
where the $\vb{Q}_n$ are a subset of the reciprocal vectors of the superlattice. In this work we will only consider triangular and square superlattice geometries, for which we take $\vb{Q}_n = 4\pi/\sqrt{3}L (\cos(2\pi n/6), \sin(2\pi n/6))$, $n=0,\ldots,5$ and $\vb{Q}_n = 2\pi/L (\cos(2\pi n/4), \sin(2\pi n/4))$, $n=0,\ldots,3$, respectively. $L$ defines the superlattice periodicity. Note also that in the experimental setup depicted in Fig.~\ref{fig:exp-setup}, $V_{\text{SL}}$ and $V_0$ can be tuned independently by varying the voltages on the unpatterned (top) and patterned (bottom) electrodes.

To summarize, the model has eight numerical parameters. For the hopping parameters in $\mathcal{H}_{\text{BLG}}$, we use values of $t = \SI{380}{meV}$, $v = \SI{673}{meV.nm}$, $v_3 = \SI{81}{meV.nm}$ and $v_4 = \SI{30}{meV.nm}$, as determined by infrared spectroscopy \cite{kuzmenko2009params}. The screening constant $\alpha$ is determined by both the material properties of graphene and the gate design, and is taken to be 0.3 based on Ref.~\cite{rokni2017layer}. That leaves $L$, $V_{\text{SL}}$, and $V_0$ as the tunable parameters. For the analysis presented in Sec.~\ref{sec:result}, we fix $L = \SI{30}{nm}$ and scan the remaining parameter space spanned by $V_{\text{SL}}$ and $V_0$. In Sec.~\ref{sec:scale} we briefly discuss how the results vary with $L$.

\section{Single-particle FCI indicators}
\label{sec:indicators}
A Bloch band is described by its dispersion $E(\vb{k})$ and its quantum geometry. 
The latter is encoded in the quantum geometric tensor (QGT), defined as
\begin{equation}
    G^{ab}(\vb{k}) = \braket{D^a u_{\vb{k}}}{D^b u_{\vb{k}}},
\end{equation}
where $D^a = \partial^a_{\vb{k}} - i A^a(\vb{k})$ is the covariant derivative in momentum space, $A^a(\vb{k}) = -i \braket{u_{\vb{k}}}{\partial^a_{\vb{k}} u_{\vb{k}}}$ is the Berry connection, and $u_{\vb{k}}$ is the periodic part of the Bloch wavefunction with quasimomentum $\vb{k}$. The QGT is a gauge-invariant quantity that captures how $u_{\vb{k}}$ varies with $\vb{k}$. Since the QGT is Hermitian, its real part, denoted $g^{ab}(\vb{k})$, is a symmetric form referred to as the quantum metric or the Fubini-Study metric. Its imaginary part is equal to $\frac{1}{2} \epsilon^{ab} \Omega(\vb{k})$, where $\epsilon^{ab}$ is the two-dimensional anti-symmetric Levi-Civita symbol and
\begin{equation}
    \Omega(\vb{k}) = \partial^1_{\vb{k}} A^2(\vb{k}) - \partial^2_{\vb{k}} A^1(\vb{k})
\end{equation}
is the (scalar) Berry curvature. The Berry curvature is related to the Chern number $\mathcal{C}$ of the band by
\begin{equation}
    \mathcal{C} = \frac{1}{2\pi} \int_{\text{BZ}} \dd{\vb{k}} \Omega(\vb{k}),
\end{equation}
where the integral is over the first Brillouin zone. The Chern number is always an integer; when $\mathcal{C} \neq 0$ we refer to the band as topological or a Chern band. At all values of $\vb{k}$, the quantum metric and Berry curvature obey the inequalities
\begin{equation} \label{tr-ineq}
    \frac{1}{2} \tr g(\vb{k}) \geq \sqrt{\det g(\vb{k})} \geq \frac{1}{2} \abs{\Omega(\vb{k})},
\end{equation}
which follows from the positive semi-definiteness of the QGT. If the Berry curvature has the same sign everywhere in the Brillouin zone, this implies
\begin{equation} \label{int-tr-ineq}
    \tr \bar{g} \geq 2\pi \abs{\mathcal{C}},
\end{equation}
where
\begin{equation} 
    \bar{g}^{ab} = \int_{\text{BZ}} \dd{\vb{k}} g^{ab}(\vb{k})
\end{equation}
is the integrated quantum metric.

In the presence of Coulomb interactions, a fractionally filled Chern band may realize an FCI ground state.
Verifying the FCI ground state computationally requires an expensive many-body calculation.
In this manuscript,  following earlier literature \cite{Parameswaran2012,Roy2014,claassen2015position,jackson2015geometric,bauer2016quantum,ledwith2020chiral,Jie2021,mera2021kahler,valentin2023,ledwith2023vortexability,morales2023pressure}, we predict the result of such a calculation using ``single-particle indicators,'' which are features of the band structure that quantify how much the Chern band resembles a lowest Landau level (LLL).
While the single-particle indicators do not guarantee an FCI ground state (and have known limitations \cite{simon2020contrasting}), they provide an efficient search of the large parameter space we want to study.

We now describe the single-particle indicators that we employ.
By imposing a magnetic Brillouin zone, the LLL can be interpreted as a Bloch-like energy band with $\abs{\mathcal{C}} = 1$, which is completely degenerate (i.e., $E(\vb{k})$ is constant), has uniform quantum geometry (i.e., $G(\vb{k})$ is constant), and saturates the inequality (\ref{tr-ineq}) \cite{Jie2021}. Based on these properties, multiple conditions for LLL mimicry have been proposed. 

The first condition is that the bandwidth
\begin{equation}
    W = \max_{\vb{k}} E(\vb{k}) - \min_{\vb{k}} E(\vb{k}) 
\end{equation}
should be minimized so that correlation effects dominate the kinetic energy contribution. 

Since the LLL has uniform Berry curvature, the second condition is that spatial variations in the Berry curvature should be minimized, i.e.,
\begin{equation}
    F = \qty[ A_{\text{BZ}} \int_{\text{BZ}} \dd{\vb{k}} \qty( \frac{\Omega(\vb{k})}{2\pi} - \frac{\mathcal{C}}{A_{\text{BZ}}})^2 ]^{1/2}
\end{equation}
should be minimized, where $A_{\text{BZ}}$ is the area of the Brillouin zone. This condition was supported by early exact diagonalization studies in a variety of lattice models \cite{jackson2015geometric}, and it was shown that in the limit of $F = 0$, the band recovers the Girvin-Macdonald-Platzman density operator algebra of the LLL at long wavelengths \cite{Parameswaran2012}.

More recently, the importance of the quantum metric and the inequality (\ref{tr-ineq}) has also been highlighted. 
Specifically, it has been shown that wavefunctions of ``ideal'' flat bands, defined as $\abs{\mathcal{C}} = 1$ bands in which $\Omega(\vb{k})$ does not vanish and (\ref{tr-ineq}) is saturated, are proportional to LLL wavefunctions with a momentum-independent prefactor \cite{Jie2021}. 
This mapping allows the construction of Laughlin-like many-body wavefunctions that are exact zero-energy eigenstates of a generalized Haldane pseudopotential Hamiltonian \cite{Jie2021}. Models that exactly satisfy these criteria do exist, e.g., the chiral limit of twisted bilayer graphene \cite{ledwith2020chiral}, though the model we consider in this work has no such known limit. Therefore, we instead seek to minimize
\begin{equation}
    \overline{T} = \abs{\tr \bar{g}} - 2\pi \abs{\mathcal{C}},
\end{equation}
which quantifies the deviation from saturation of (\ref{tr-ineq}). (Note that $\overline{T} = 0$ implies that (\ref{int-tr-ineq}) is saturated, which in turn implies saturation of (\ref{tr-ineq}) at all $\vb{k}$.) 

Although not strictly within the framework of LLL mimicry, it is also important for the band gap
\begin{multline}
    \Delta = \min\Big\{\min_{\vb{k}} E_{n+1}(\vb{k}) - \max_{\vb{k}} E_n(\vb{k}), \\ \min_{\vb{k}} E_n(\vb{k}) - \max_{\vb{k}} E_{n-1}(\vb{k})\Big\}
\end{multline}
(where we have momentarily introduced a band index to $E$, with $n$ corresponding to the band of interest) to be large compared to the interaction scale, so that the effect of neighboring bands can be ignored.

In summary, the single-particle features we seek as indicators that a band with $\abs{\mathcal{C}} = 1$ is most suitable to host a stable FCI phase are minimal $W$, $\overline{T}$, and $F$, and maximal $\Delta$.

\section{Optimizing the potential parameters}
\label{sec:result}

In this section, we compute the single-particle indicators defined in Sec.~\ref{sec:indicators} as functions of the electrically tunable parameters $V_{\text{SL}}$ and $V_0$ for multiple bands in both triangular and square sBLG. The superlattice length scale is fixed at $L = \SI{30}{nm}$, with discussion of other values deferred to Sec.~\ref{sec:scale}. The goal is to map the topological phase diagram of the system and identify the parameter regions where an FCI ground state is most likely to exist.

\subsection{Triangular superlattice}

\begin{figure}
    \centering
    \subfigure[]{\includegraphics[width=0.495\linewidth]{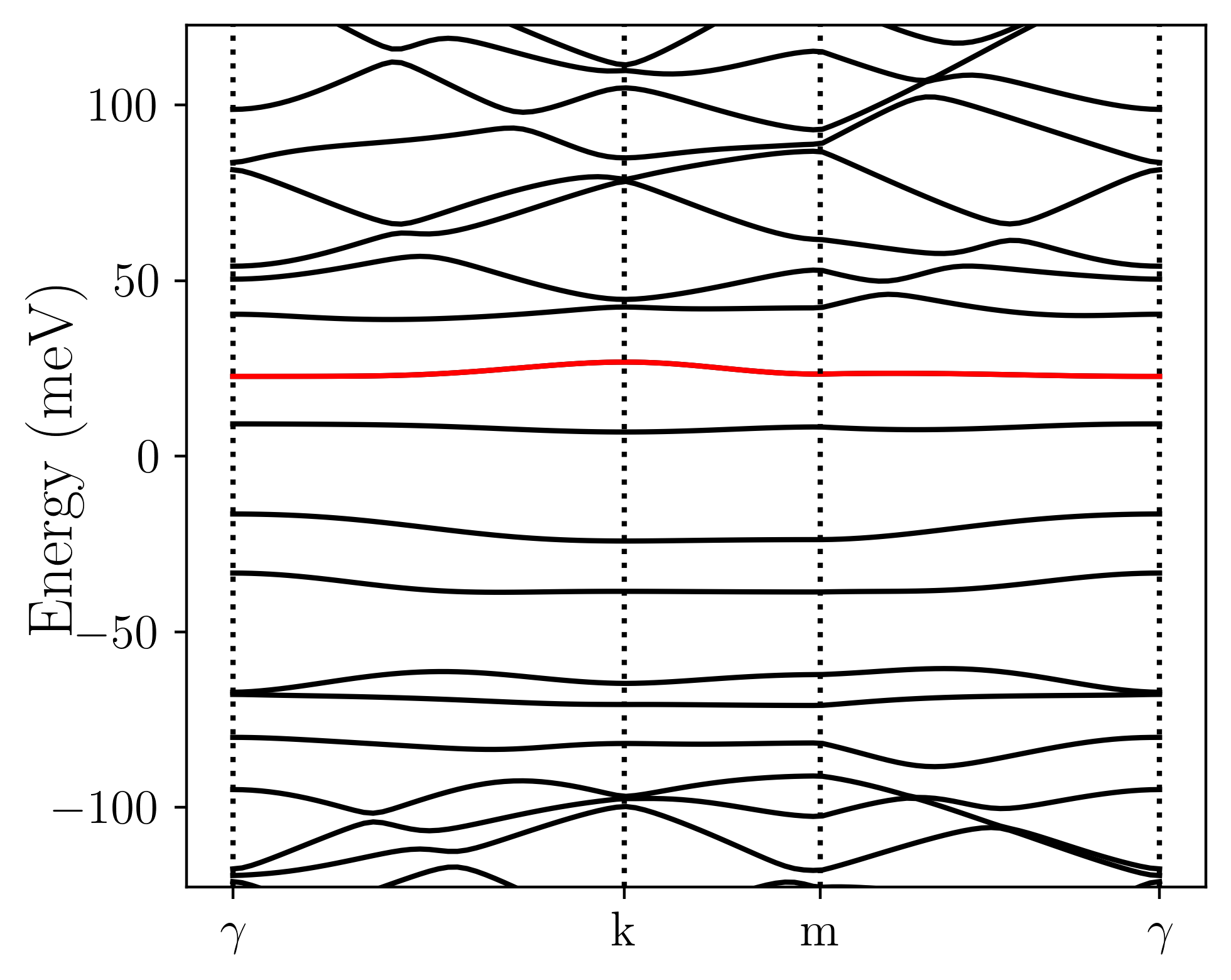}}
    \subfigure[]{\includegraphics[width=0.485\linewidth]{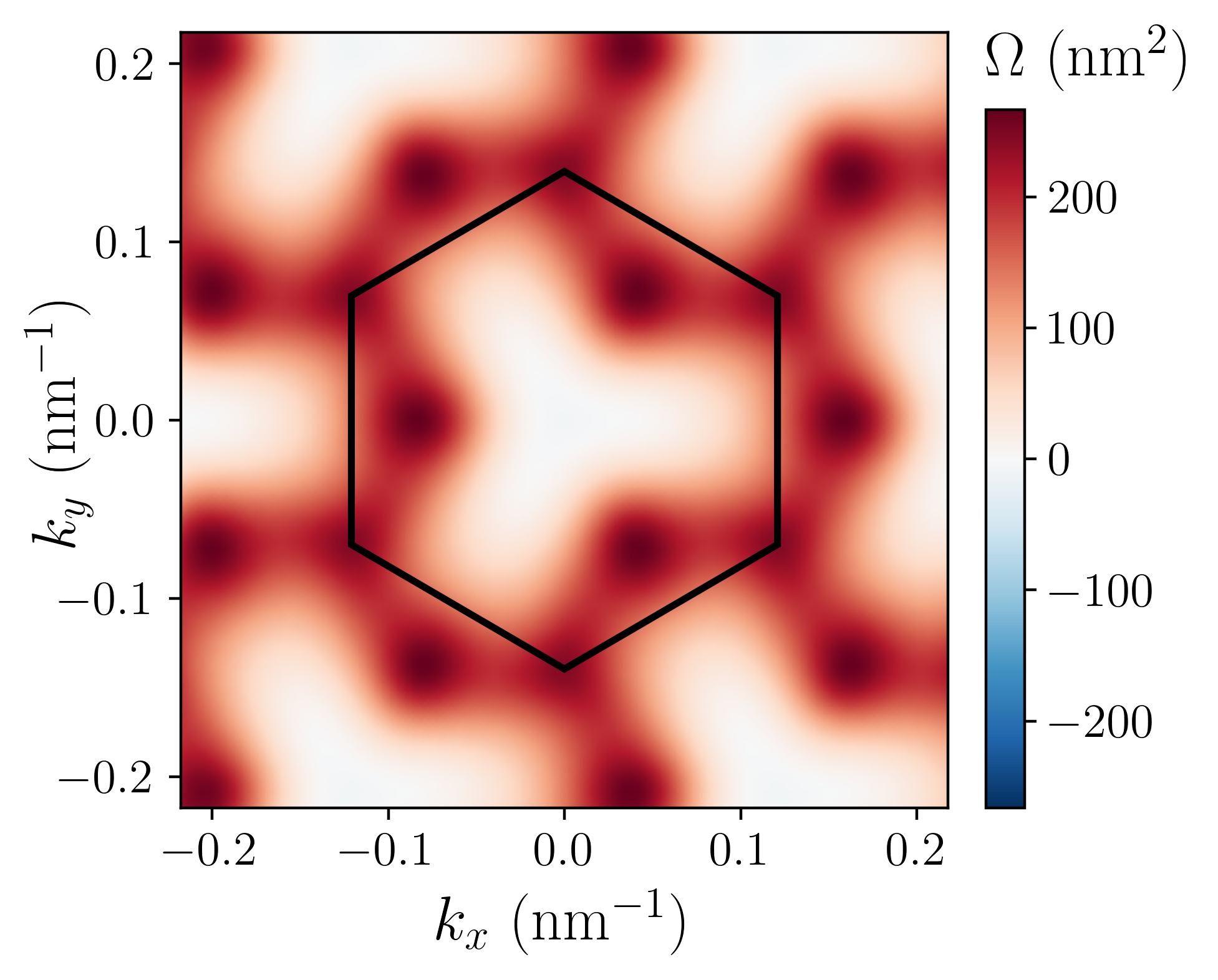}}
    \caption{(a) Band structure of triangular sBLG with $L = \SI{30}{nm}$, $V_{\text{SL}} = \SI{30}{meV}$, and $V_0 = -\SI{30}{meV}$. The $n=0$ band is highlighted. Lowercase labels on the horizontal axis refer to the high-symmetry points of the superlattice Brillouin zone. (b) Berry curvature distribution for the $n=0$ band, which is topological with $\mathcal{C} = +1$. The boundary of the first superlattice Brillouin zone is overlayed in black.}
    \label{fig:triag-bands}
\end{figure}

We begin by considering a triangular superlattice geometry with $L = \SI{30}{nm}$. 
The reciprocal lattice vectors are listed below Eq.~(\ref{eq:VSL}).
The band structure for this system is shown in Fig.~\ref{fig:triag-bands}(a) at example parameter values of $(V_{\text{SL}}, V_0) = (\SI{30}{meV}, \SI{-30}{meV})$. The appearance of multiple stacked, low-dispersion bands is typical. We label the bands in ascending energy order with an index $n$, with the $n = 0$ band highlighted in red. At $(V_{\text{SL}}, V_0) = (0,0)$, the $n = 0$ band is the lowest band lying above the Fermi level, but at the parameter values shown it is the second lowest. Furthermore, at these parameter values, the $n = 0$ band is topological with $\mathcal{C} = +1$ (see Fig.~\ref{fig:triag-bands}(b) for the Berry curvature distribution), while the neighboring $n = 1$, $-1$, $-2$, and $-3$ flatbands are topologically trivial. We will ultimately find that the $n = 0$ band is topological across the widest range of parameter values and hosts the most optimal indicator values, so it is the first we will consider.

\subsubsection{$n = 0$ band}

\begin{figure}
    \centering
    \subfigure[]{\includegraphics[height=0.42\linewidth]{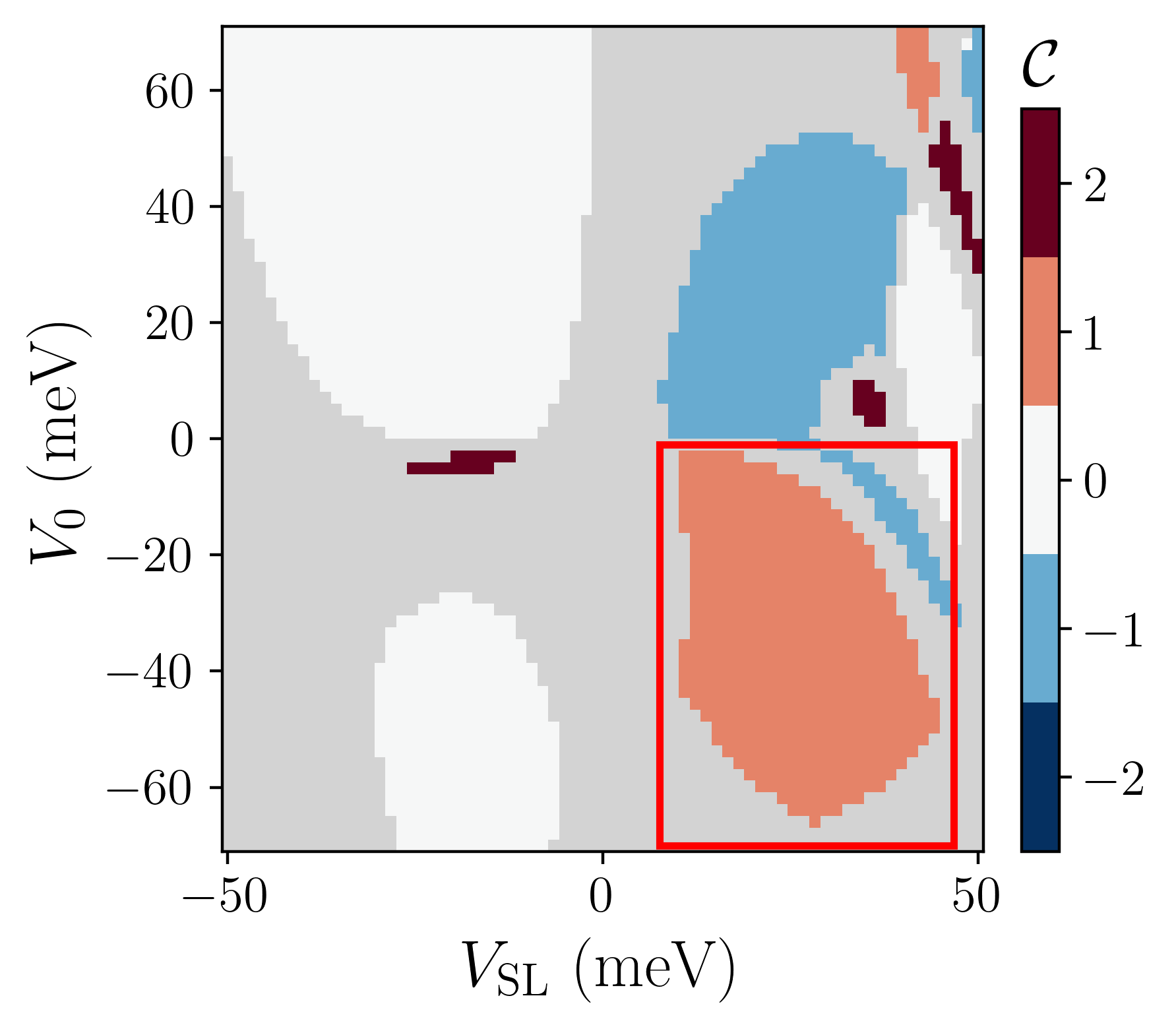}}
    \subfigure[]{\includegraphics[height=0.42\linewidth]{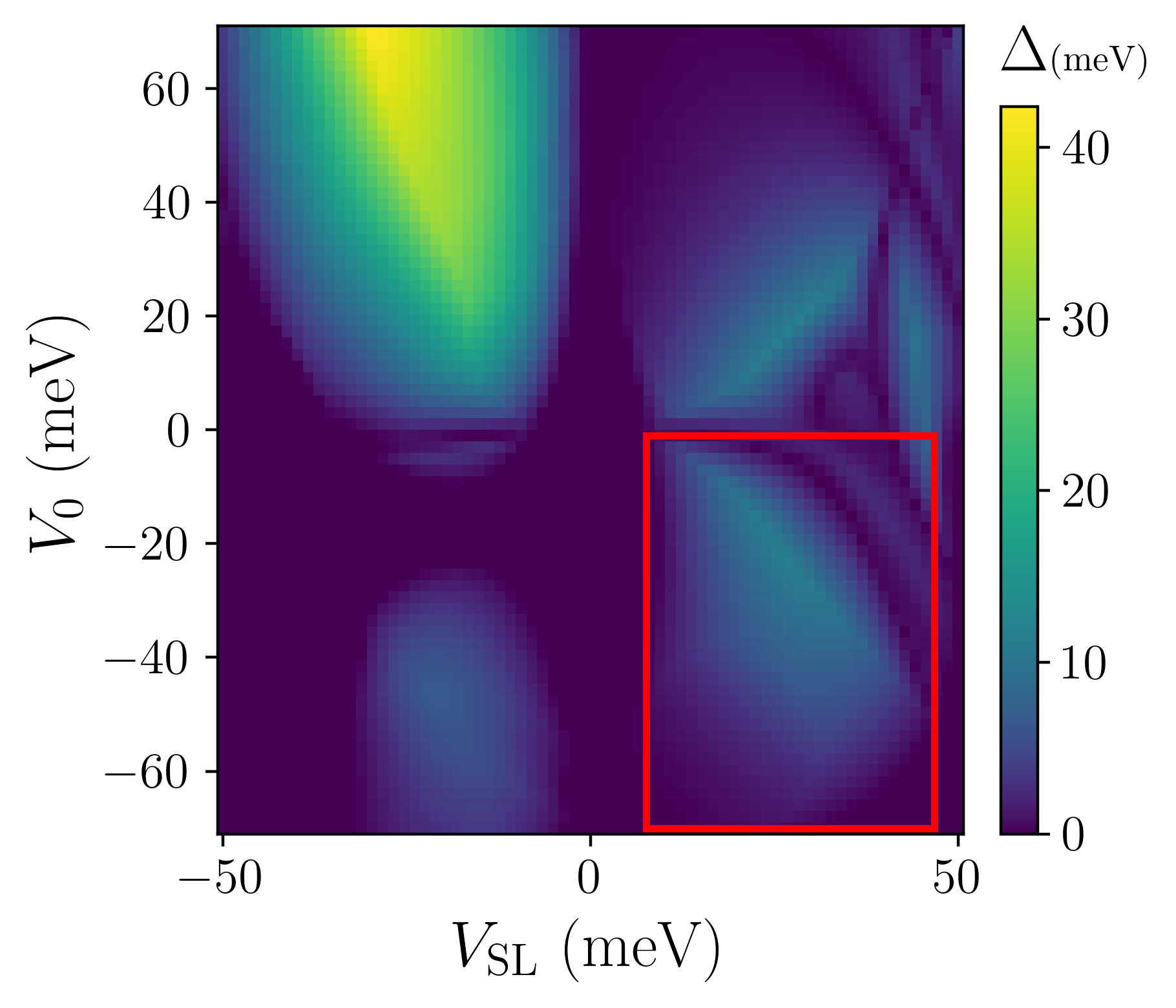}}
    \caption{(a) Chern number $\mathcal{C}$ and (b) band gap $\Delta$ of the $n=0$ band of $L = \SI{30}{nm}$ triangular sBLG. Grayed regions of (a) indicate that the Chern number is not numerically well-resolved due to the proximity of a band closing. The red box highlights the $\mathcal{C}=+1$ phase which is of primary interest for realizing an FCI. Single-particle indicators in this region are shown in Fig.~\ref{fig:triag-n0-zoom}.}
    \label{fig:triag-n0-phase}
\end{figure}

\begin{figure}
    \begin{tabular}{l l}
        \subfigure[]{\includegraphics[height=0.42\linewidth]{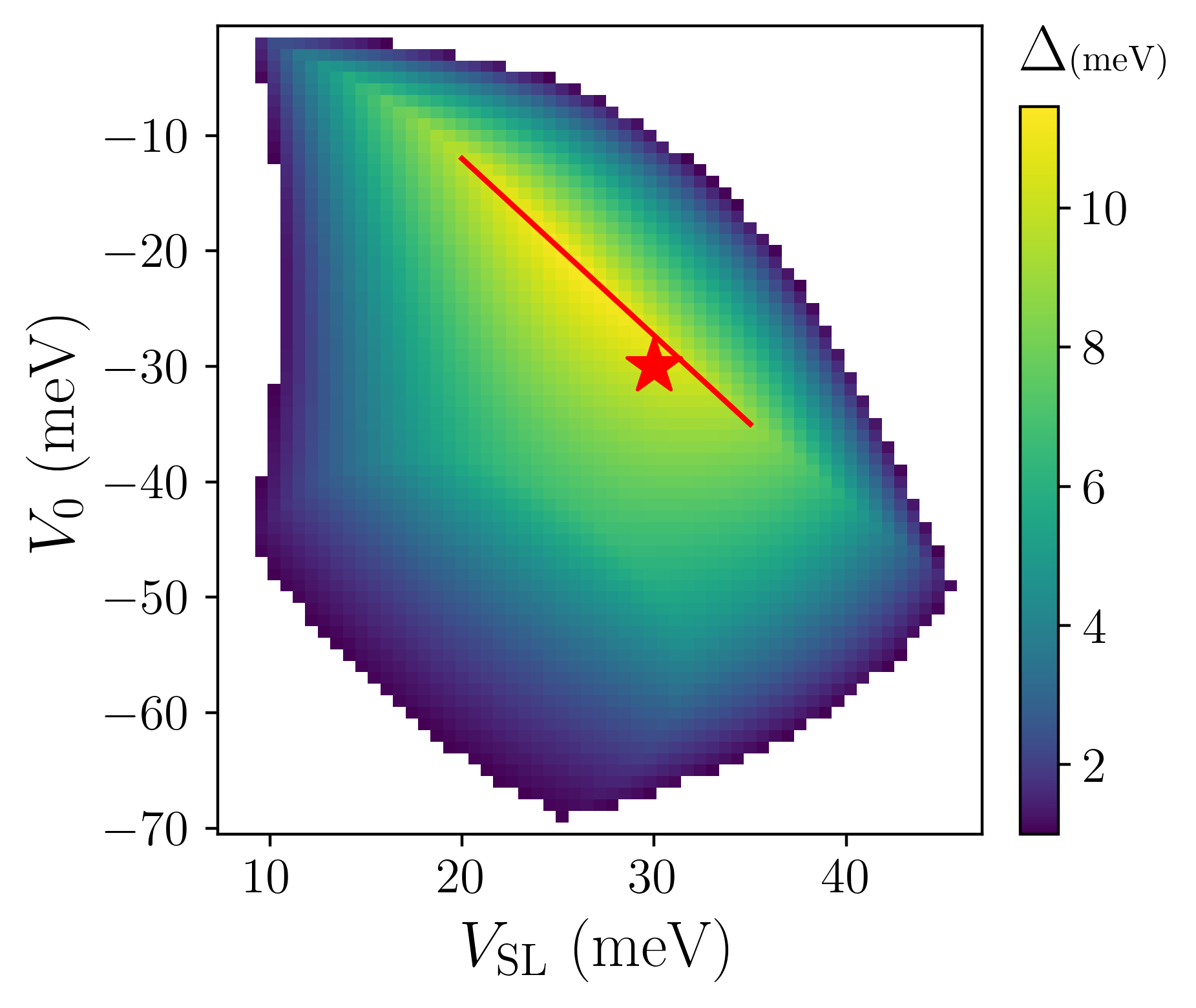}} &
        \subfigure[]{\includegraphics[height=0.42\linewidth]{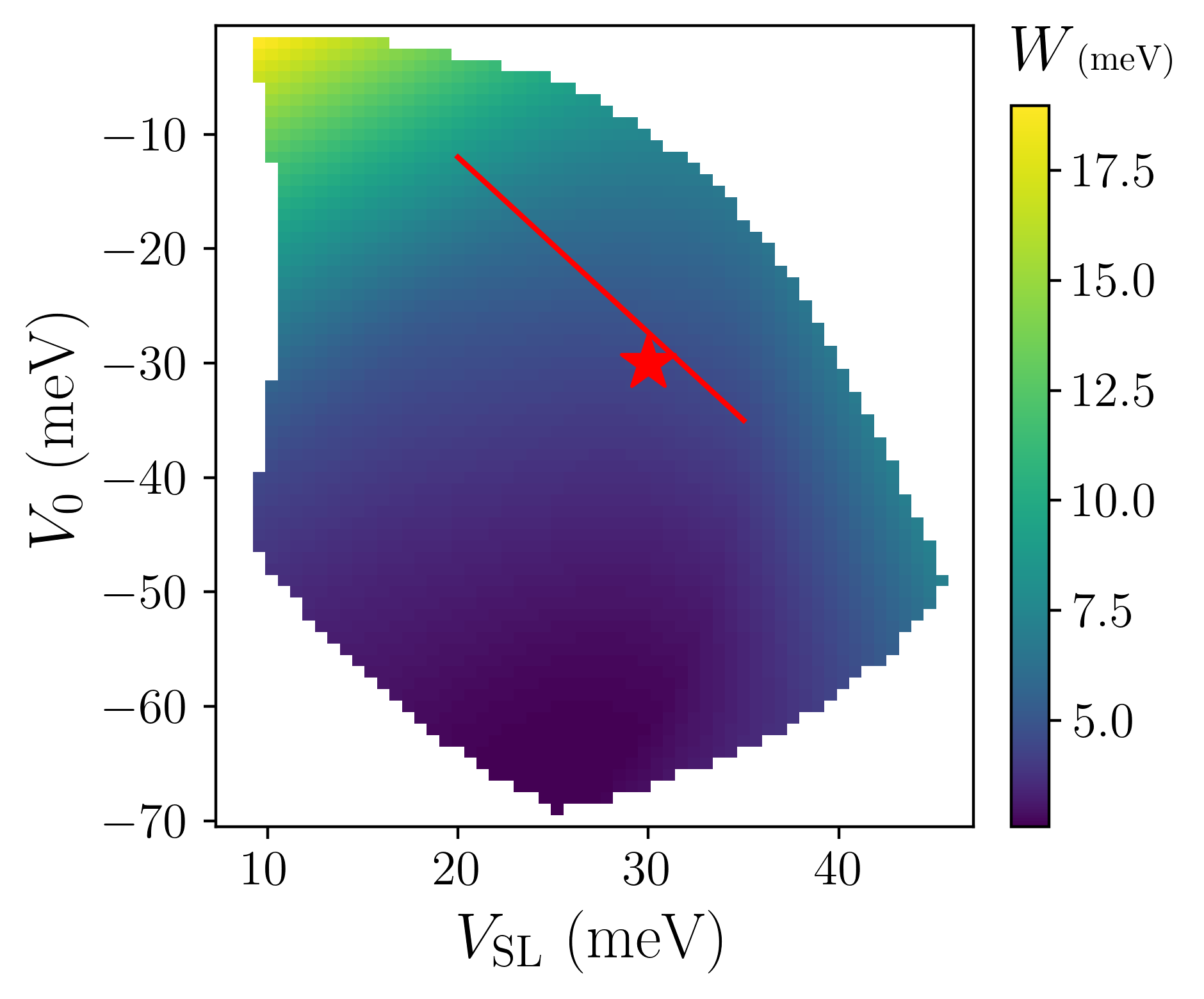}} \\
        \subfigure[]{\includegraphics[height=0.42\linewidth]{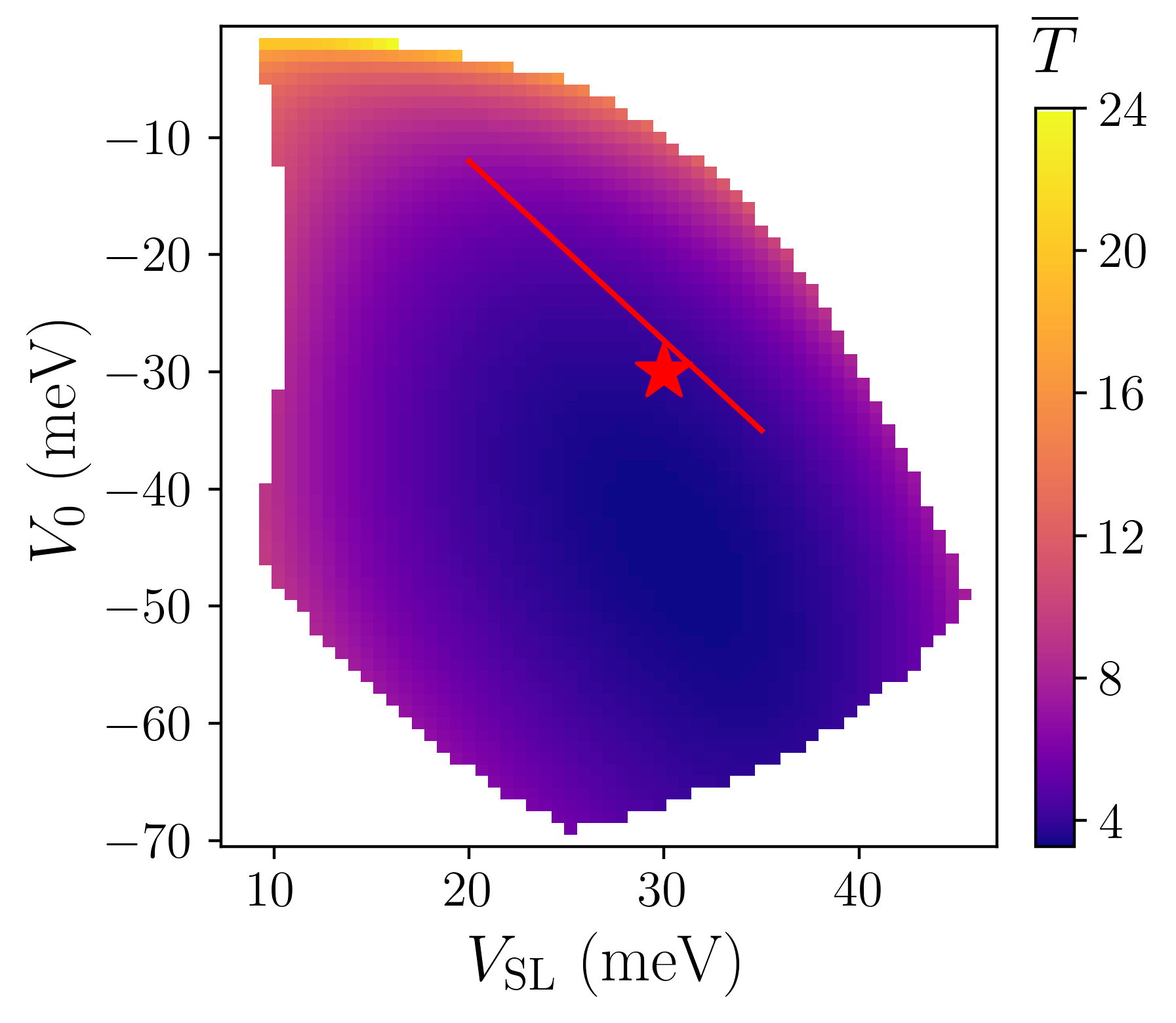}} &
        \subfigure[]{\includegraphics[height=0.42\linewidth]{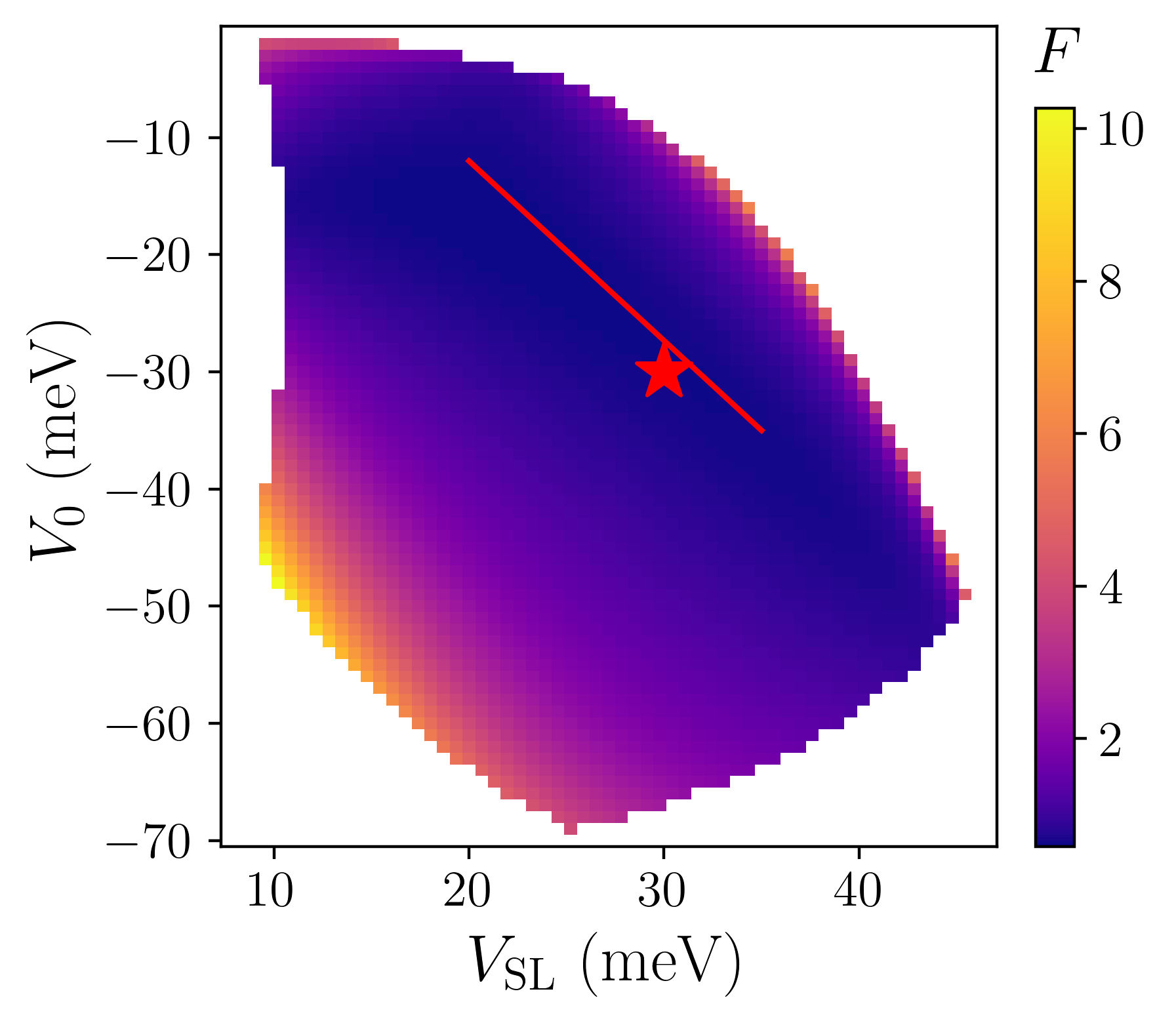}}
    \end{tabular}
    \caption{Indicators (a) $\Delta$, (b) $W$, (c) $\overline{T}$, and (d) $F$ for the $n=0$ band of triangular sBLG, in the $C = +1$ phase highlighted in Fig.~\ref{fig:triag-n0-phase}. 
    The ``optimal line'' along which $\Delta$ and $F$ are jointly optimized is indicated.
    The red star marks $(V_{\text{SL}}, V_0) = (\SI{30}{meV}, -\SI{30}{meV})$, the parameters used in Fig.~\ref{fig:triag-bands}, which the indicators shown here also suggest are near-optimal for realizing an FCI in this region.}
    \label{fig:triag-n0-zoom}
\end{figure}

Fig.~\ref{fig:triag-n0-phase} shows the Chern number $\mathcal{C}$ and band gap $\Delta$ of the $n = 0$ band as functions of $V_{\text{SL}}$ and $V_0$. (Gray regions of Fig.~\ref{fig:triag-n0-phase}(a) indicate that the Chern number is not numerically well-resolved due to the proximity of a band closing; note the coincidence of these regions with dark regions of Fig.~\ref{fig:triag-n0-phase}, where $\Delta$ is small.) There are two prominent, approximately symmetric $\abs{\mathcal{C}} = 1$ phases present which are of primary interest in our search for a stable FCI state. (When $\alpha = 1$, the two phases are related by an exact symmetry under $PT$, where $P$ is inversion around an AB stacking point and $T$ is time reversal. $PT$ commutes with $H_{\text{BLG}}$ but sends $(V_{\text{SL}}, V_0) \to (V_{\text{SL}}, -V_0)$.) 
The $\mathcal{C} = +1$ region with $V_0 <~0$ (indicated by the red box) has slightly more favorable indicators and so will be our focus. Other topological phases are present as well, including some with $\abs{\mathcal{C}} > 1$; however, these phases require fine-tuning of the parameters $(V_{\text{SL}}, V_0)$ to within a few meV and do not attain band gaps of greater than 5 meV, so we do not consider them in this work.

\begin{figure}
    \begin{tabular}{| r | c | c | c | c | c |}
        \hline & & & & \\[-1em]
        & $\Delta$ (meV) & $W$ (meV) & $U_c$ (meV) & $\overline{T}$ & $F$ \\[0.1em] \hline
        sBLG & 10.3 & 4.5 & $\sim 5$ & 3.8 & 0.63 \\
        TBG & 12 & 11 & $\sim 20$ & 2-5 & 2-12 \\
        TBG+HF & 0 & 30-40 & $\sim 20$ & 2-4 & 2-3 \\
        tMoTe\textsubscript{2} & 8 & 6.5 & $\sim 30$ & 1 & 0.1 \\
        tMoTe\textsubscript{2}+HF & 35 & 1.2 & $\sim 30$ & 0.7 & 0.2 \\ \hline
    \end{tabular}
    \caption{Comparison of FCI indicators for $L=\SI{30}{nm}$ triangular sBLG at $(V_{\text{SL}}, V_0) = (\SI{30}{meV}, -\SI{30}{meV})$ in the $n = 0$ band; $\theta = 1.05\degree$ magic-angle TBG, both with \cite{parker2021} and without \cite{nam2017tbg,ledwith2020chiral} Hartree-Fock (HF) corrections; and $\theta = 3.7\degree$ twisted MoTe\textsubscript{2} (tMoTe\textsubscript{2}) in the top valence band, both with and without HF corrections \cite{dong2023}. Values for TBG are highly sensitive to the ``chiral ratio'' $\kappa$, the physical value of which is not known precisely. Quoted ranges correspond to $\kappa$ between 0.6 and 0.8, with the exception of $\Delta$ and $W$ for uncorrected TBG, which was instead computed from a first-principles model of lattice relaxation \cite{nam2017tbg}. The third column lists an extremely coarse estimate of the Coulomb interaction scale for the sake of comparison with $\Delta$ and $W$, given by $U_c \sim e^2/4\pi\epsilon L$ with $\epsilon/\epsilon_0 \sim 10$.}
    \label{fig:ind-table}
\end{figure}

The single-particle indicators for FCI stability---band gap $\Delta$,  band width $W$, idealness deviation $\overline{T}$, and Berry curvature flucutuation $F$---are shown for the $\mathcal{C} = +1$ phase in Fig.~\ref{fig:triag-n0-zoom}. $\Delta$ is sharply maximized in a linear region of parameter space, extending from roughly $(V_{\text{SL}}, V_0) = (\SI{20}{meV}, -\SI{12}{meV})$ to $(\SI{35}{meV}, -\SI{35}{meV})$. (See the red line in Fig.~\ref{fig:triag-n0-zoom}.) $F$ is also minimized, though less sharply, in the same region, which is consistent with the tendency for Berry curvature to accumulate in the Brillouin zone near band closings. We therefore refer to this region as the ``optimal line.'' In comparison, $W$ and $\overline{T}$ are both relatively uniform over most of the region, but tend to gradually decrease as $V_{\text{SL}}$ becomes more negative. We thus conclude that the optimal parameters for realizing an FCI ground state in this phase, as indicated by single-particle energetics and band geometry, will likely lie near the lower part of the optimal line. For example, at $(V_{\text{SL}}, V_0) = (\SI{30}{meV}, -\SI{30}{meV})$ (the same parameters used for Fig.~\ref{fig:triag-bands}, also indicated by a red star in Fig.~\ref{fig:triag-n0-zoom}), we find $\Delta = \SI{10.3}{meV}$, $W = \SI{4.5}{meV}$, $\overline{T} = 3.8$, and $F = 0.63$. This is tabulated in Fig.~\ref{fig:ind-table} together with estimates for the same indicators in magic angle TBG and twisted MoTe\textsubscript{2} for comparison. 
Since the energetic indicators $\Delta$ and $W$ should rightfully be compared against the Coulomb interaction scale $U_c$, which is expected to differ between these systems due to their different superlattice periodicities, a crude estimate of $U_c \sim e^2/4\pi\epsilon L$ is also included in the table.

We finally remark that descending further below the optimal line (i.e. making $V_0$ more negative) can lead to modest improvements in $W$ and $\overline{T}$ at the cost of more significant penalties in $\Delta$ and $F$. For example, at  $(V_{\text{SL}}, V_0) = (\SI{30}{meV}, -\SI{45}{meV})$, the indicator values are $\Delta = \SI{6.6}{meV}$, $W = \SI{3.3}{meV}$, $\overline{T} = 3.3$, and $F = 0.93$. Whether this turns out to be more favorable overall for realizing the FCI depends on the relative significance of the indicators.

\subsubsection{$n = 1$ band}

\begin{figure}
    \centering
    \subfigure[]{\includegraphics[height=0.42\linewidth]{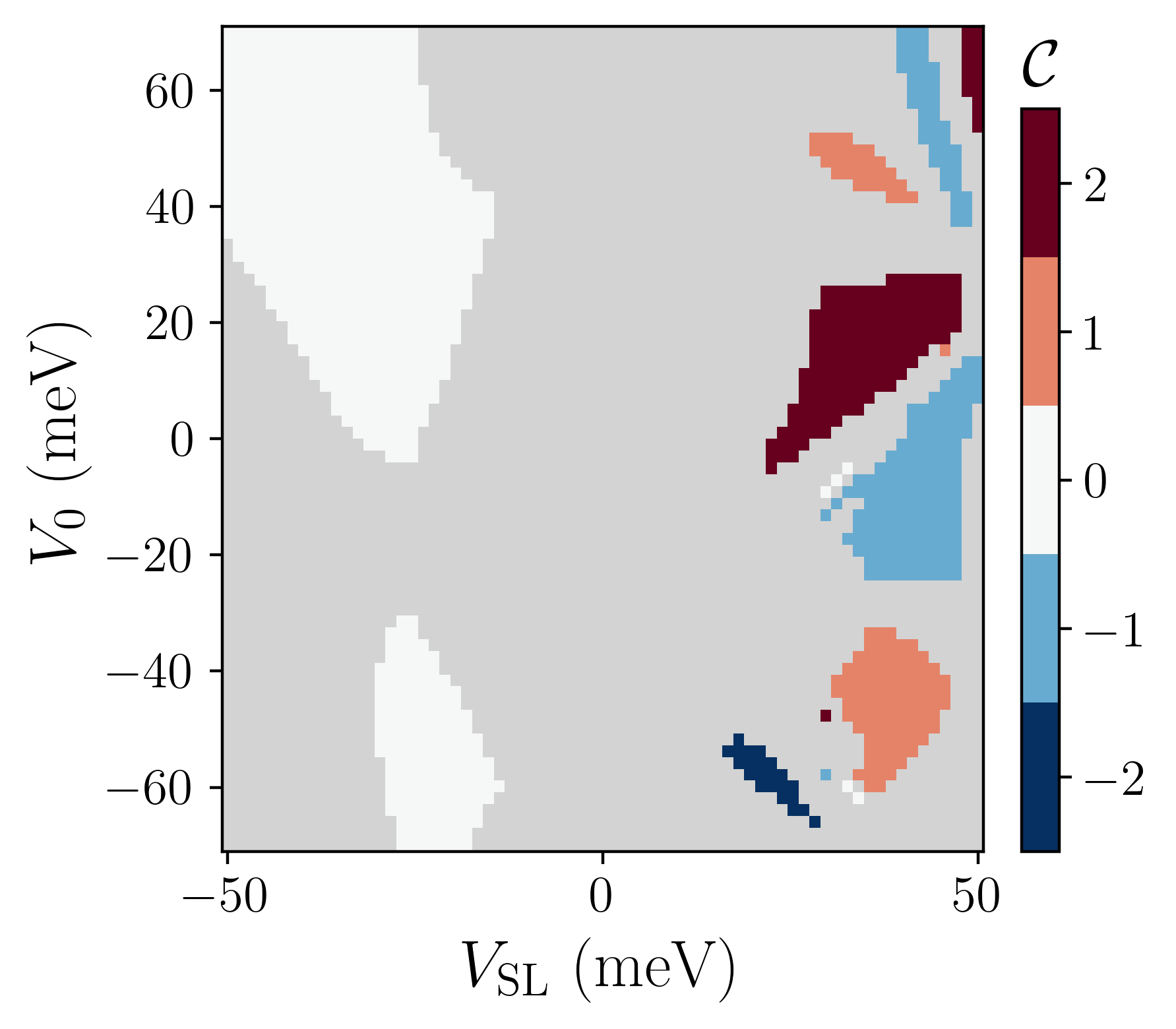}}
    \subfigure[]{\includegraphics[height=0.42\linewidth]{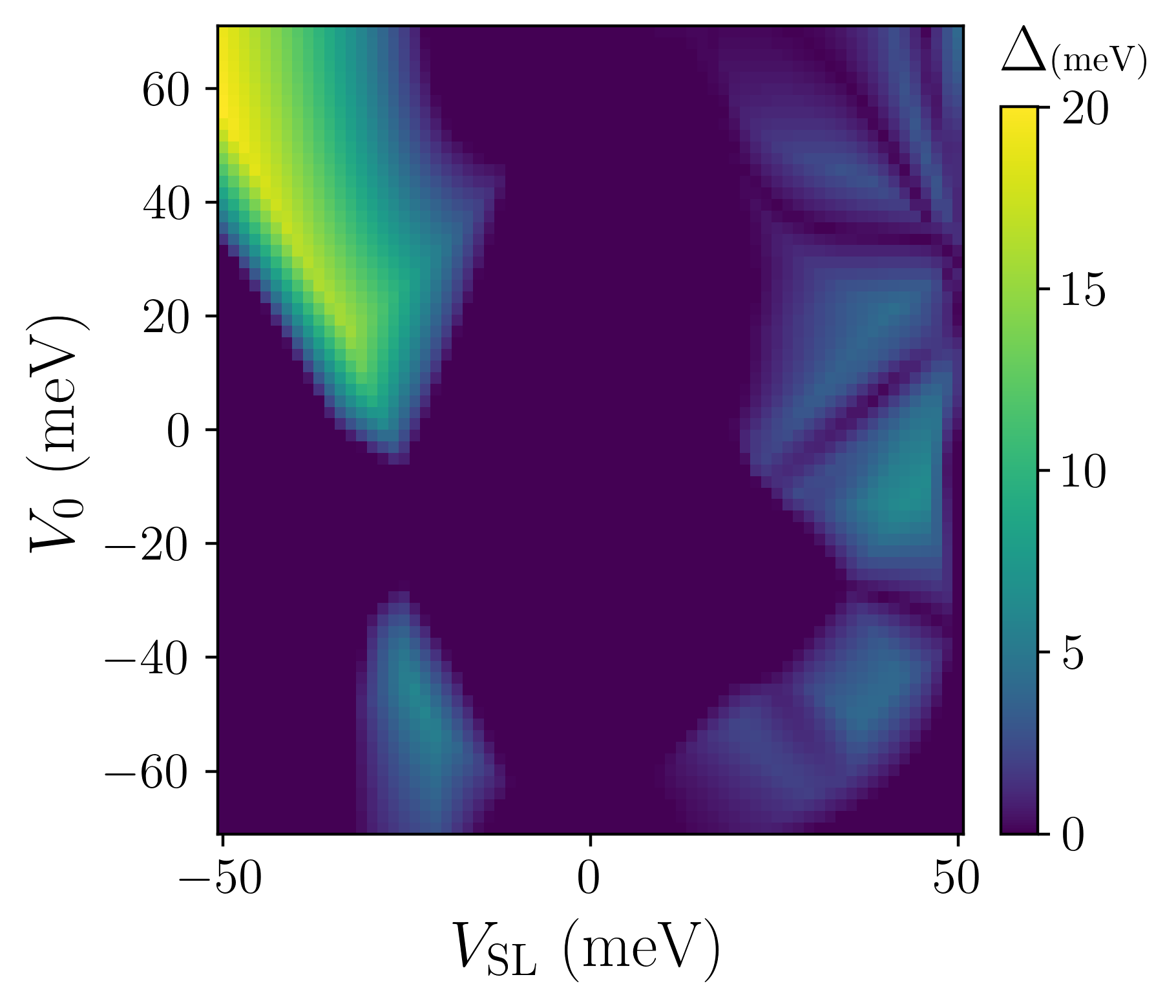}}
    \caption{(a) Chern number $\mathcal{C}$ and (b) band gap $\Delta$ of the $n=1$ band of $L = \SI{30}{nm}$ triangular sBLG. Grayed regions of (a) indicated that the Chern number is not numerically well-resolved due to the proximity of a band closing.}
    \label{fig:triag-n1-phase}
\end{figure}

$\mathcal{C}$ and $\Delta$ for the $n = 1$ band as functions of $V_\text{SL}$ and $V_0$ are shown in Fig.~\ref{fig:triag-n1-phase}. There are three relatively prominent topological phases, with $\mathcal{C} = +2$, $-1$, and $+1$, in descending order of the $V_0$ at which they appear. 
Of these, the $\mathcal{C} = -1$ phase has the most favorable indicators, attaining optimal values of $\Delta = \SI{5.9}{meV}$, $W = \SI{8.4}{meV}$, $\overline{T} = 8.6$, and $F = 1.4$ at $(V_{\text{SL}}, V_0) = (\SI{43}{meV}, \SI{-11}{meV})$. These are significantly worse than the optimal values found in the $n = 0$ band, so we conclude that the $n = 1$ band is not a good candidate in which to search for an FCI.
However, we do remark that the $\mathcal{C} = +2$ region exhibits the largest band gap ($\Delta = \SI{4.3}{meV}$) and range of parameter values of any $\abs{\mathcal{C}} > 1$ phase in a studied band and thus offers the most promising path to realizing higher-Chern physics in this system. 

\subsubsection{$n < 0$ bands}

There are flat bands with index $n < 0$ which can also become topological in certain parameter regimes. However, the phase diagrams of these bands are related to those of bands with $n \geq 0$ by a symmetry.
If $v_4 = 0$ then $H_{\text{BLG}}$ has an electron-hole symmetry generated by $\sigma_z$ in the sublattice basis, i.e., $\sigma_z$ anticommutes with $\mathcal{H}_{\text{BLG}}$. Since $\sigma_z$ commutes with $H_V$, conjugation by $\sigma_z$ together with an inversion of the parameters $(V_{\text{SL}}, V_0) \to (-V_{\text{SL}}, -V_0)$ leads to an overall inversion $H \to -H$. Thus, each $n < 0$ band is mapped by $\sigma_z$ to an $n \geq 0$ band (specifically, $n \to -1 - n$) with a phase diagram that is identical upon inversion of $V_{\text{SL}}$ and $V_0$.
Since $v_4$ is small, this correspondence is only slightly broken, so we will not comment further on the $n < 0$ bands.

\subsection{Square superlattice}

\begin{figure}
    \centering
    \subfigure[]{\includegraphics[width=0.495\linewidth]{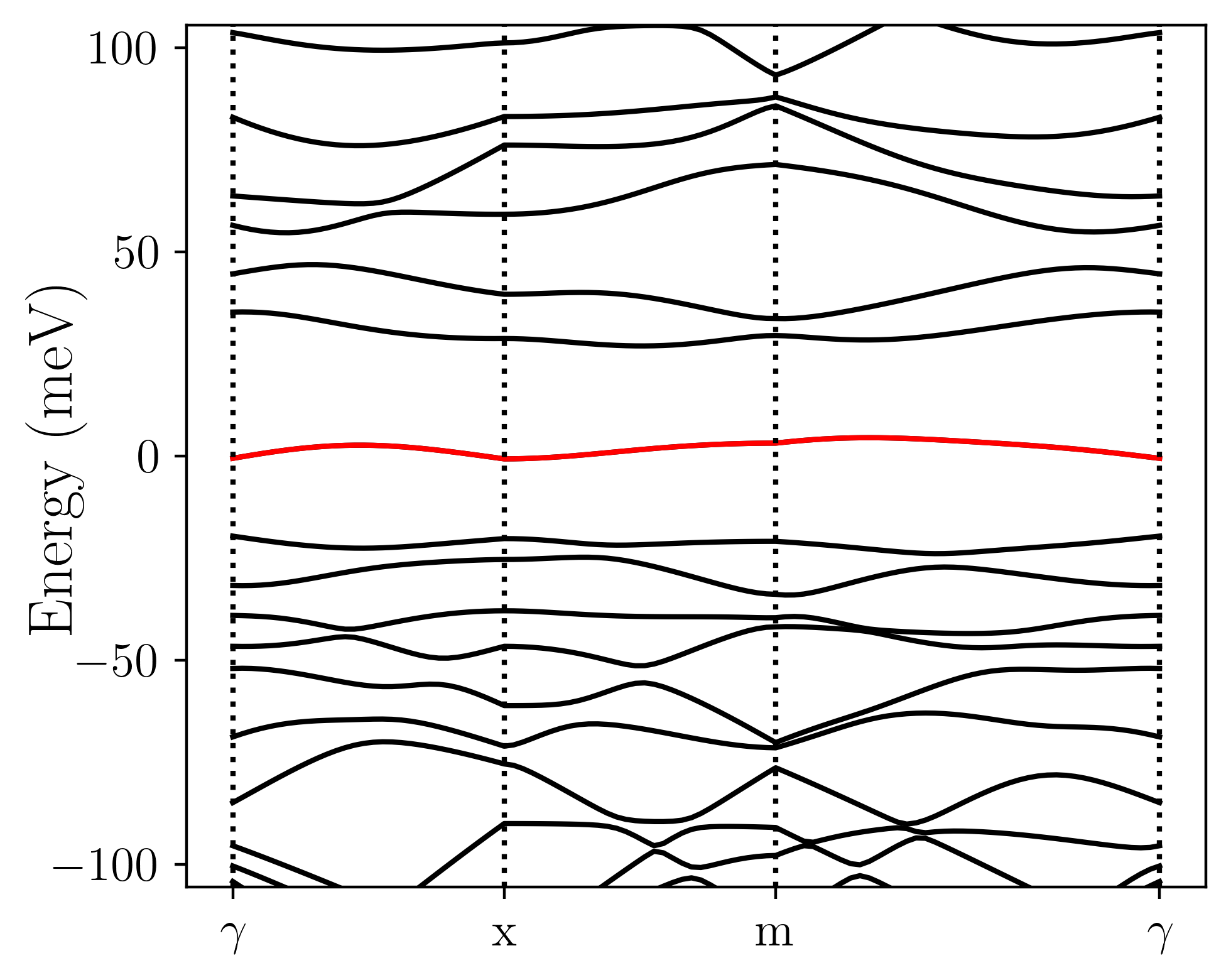}}
    \subfigure[]{\includegraphics[width=0.485\linewidth]{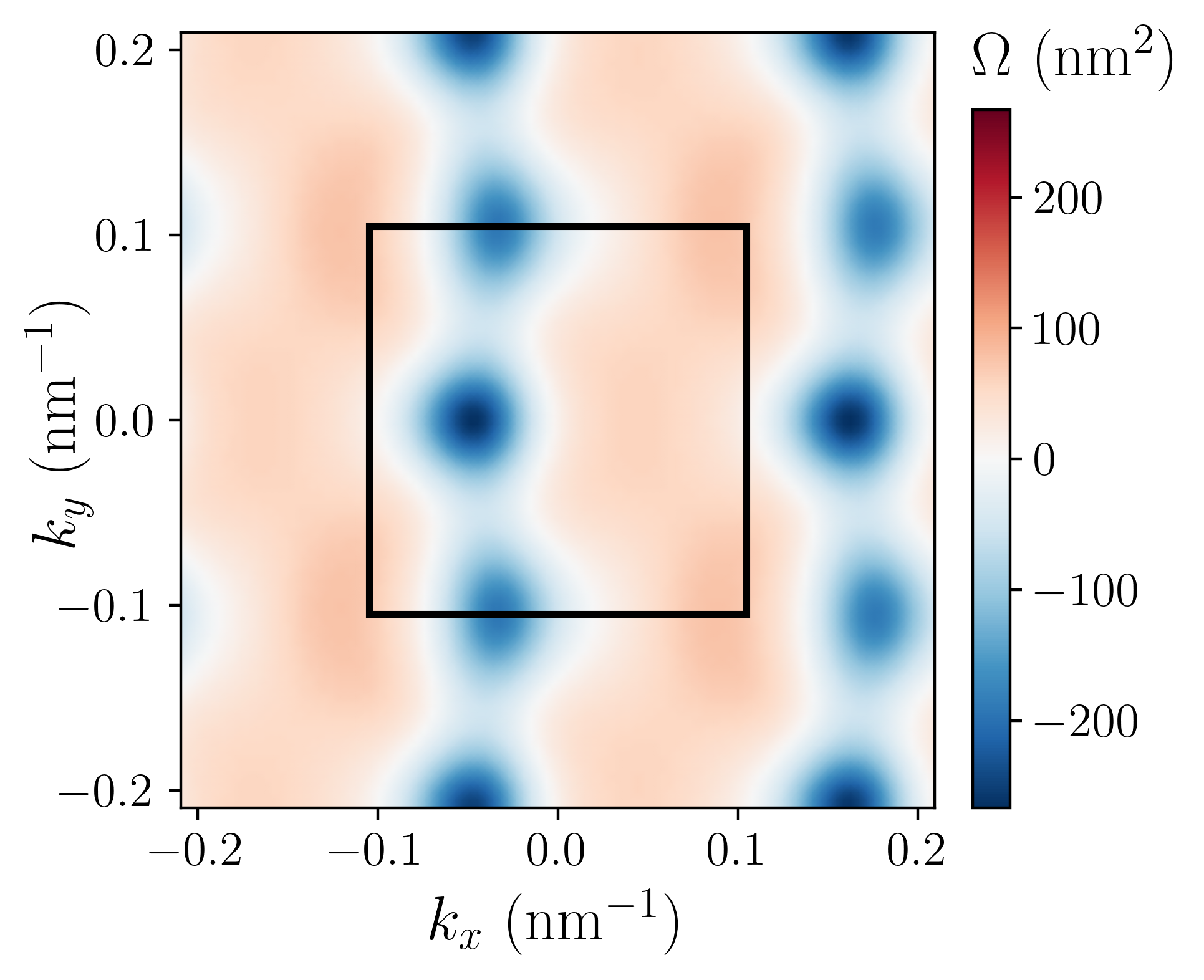}}
    \caption{(a) Band structure of square sBLG with $L = \SI{30}{nm}$, $V_{\text{SL}} = \SI{30}{meV}$, and $V_0 = \SI{30}{meV}$. The $n = 0$ band is highlighted. (b) Berry curvature distribution for the $n = 0$ band, which is topologically trivial. The boundary of the first superlattice Brillouin zone is overlayed in black.}
    \label{fig:sq-bands}
\end{figure}

\begin{figure}
    \centering
    \subfigure[]{\includegraphics[height=0.42\linewidth]{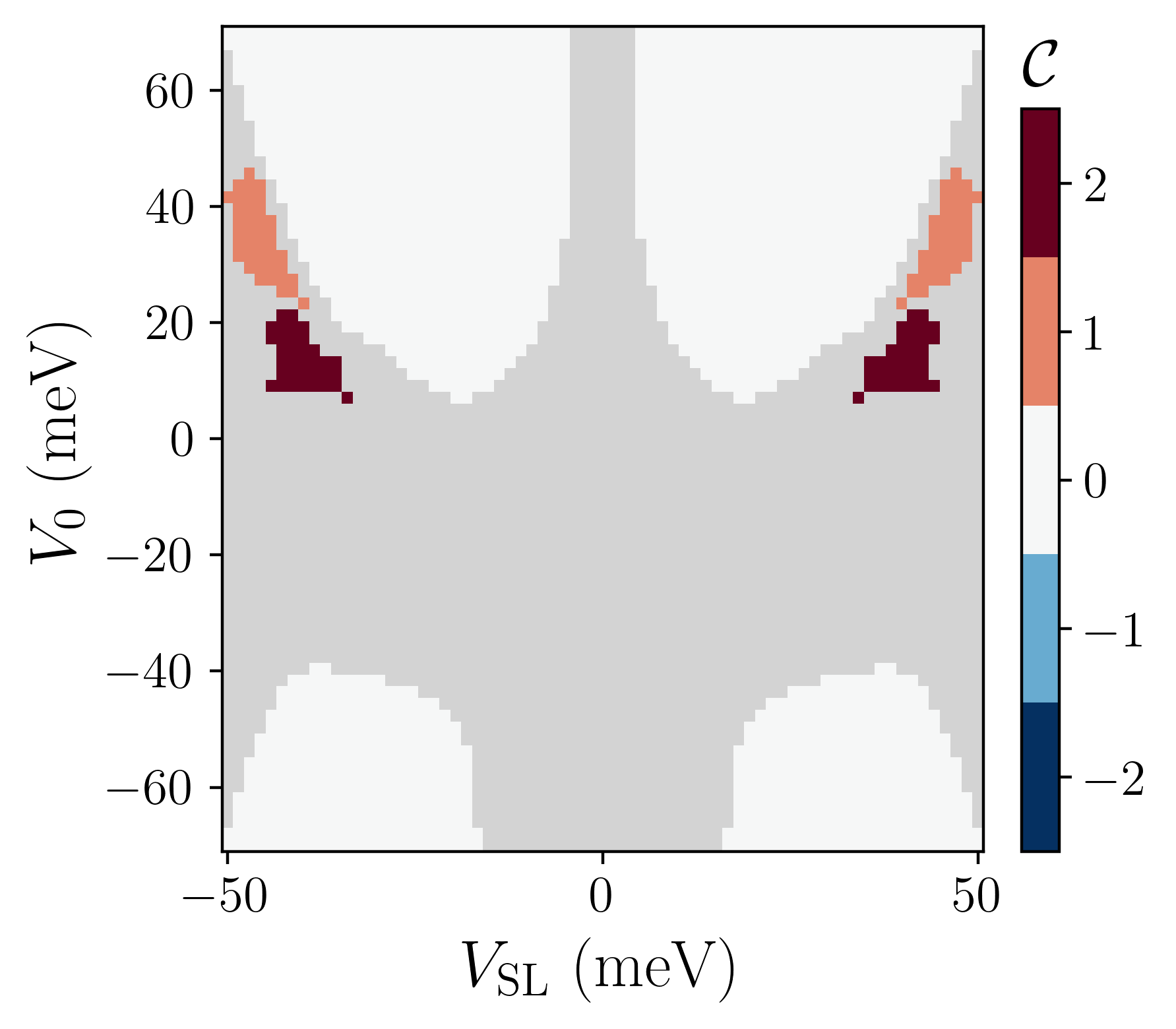}}
    \subfigure[]{\includegraphics[height=0.42\linewidth]{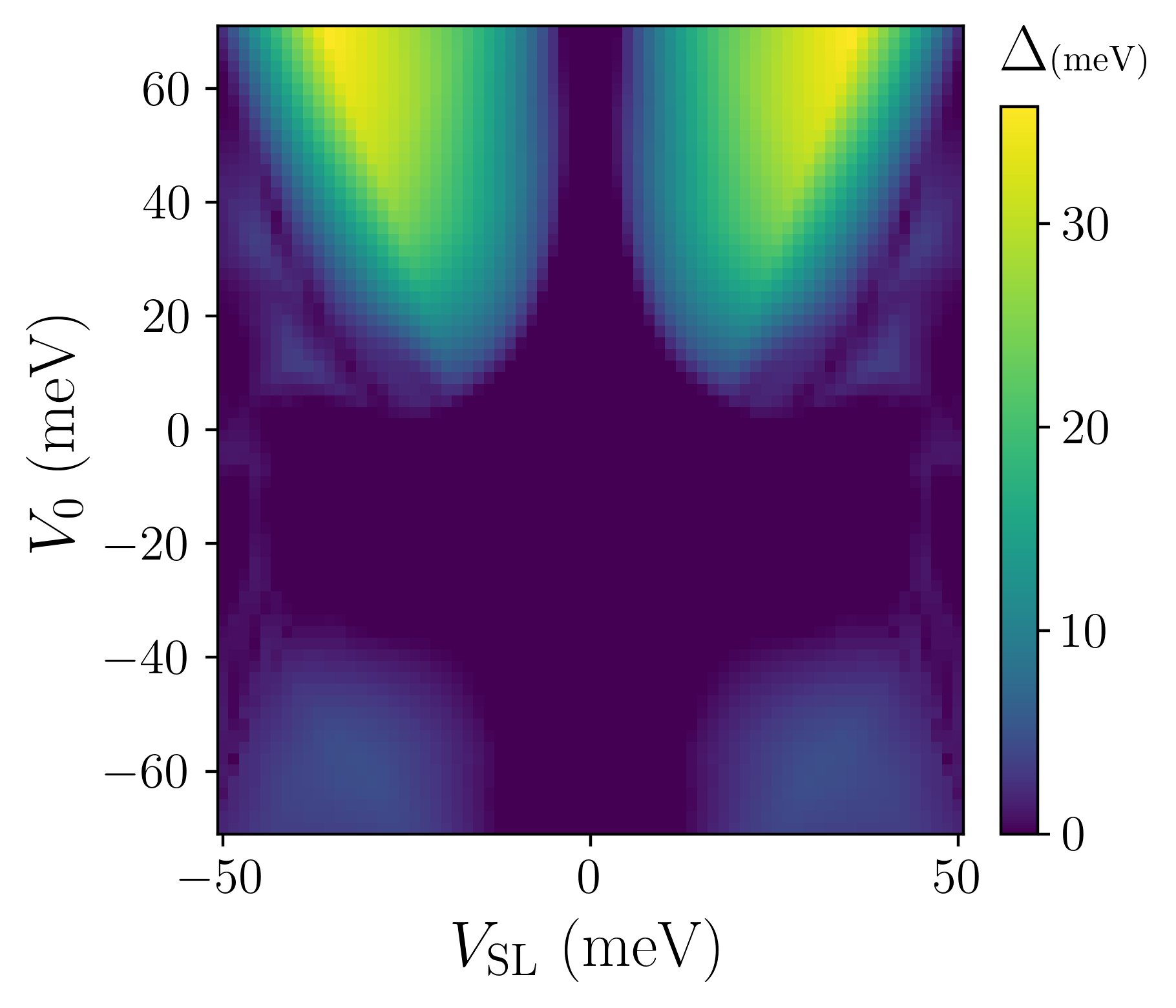}}
    \caption{(a) Chern number $\mathcal{C}$ and (b) band gap $\Delta$ of the $n=0$ band of $L = \SI{30}{nm}$ square sBLG. Grayed regions of (a) indicated that the Chern number is not numerically well-resolved due to the proximity of a band closing.}
    \label{fig:sq-n0-phase}
\end{figure}

\begin{figure}
    \centering
    \subfigure[]{\includegraphics[height=0.42\linewidth]{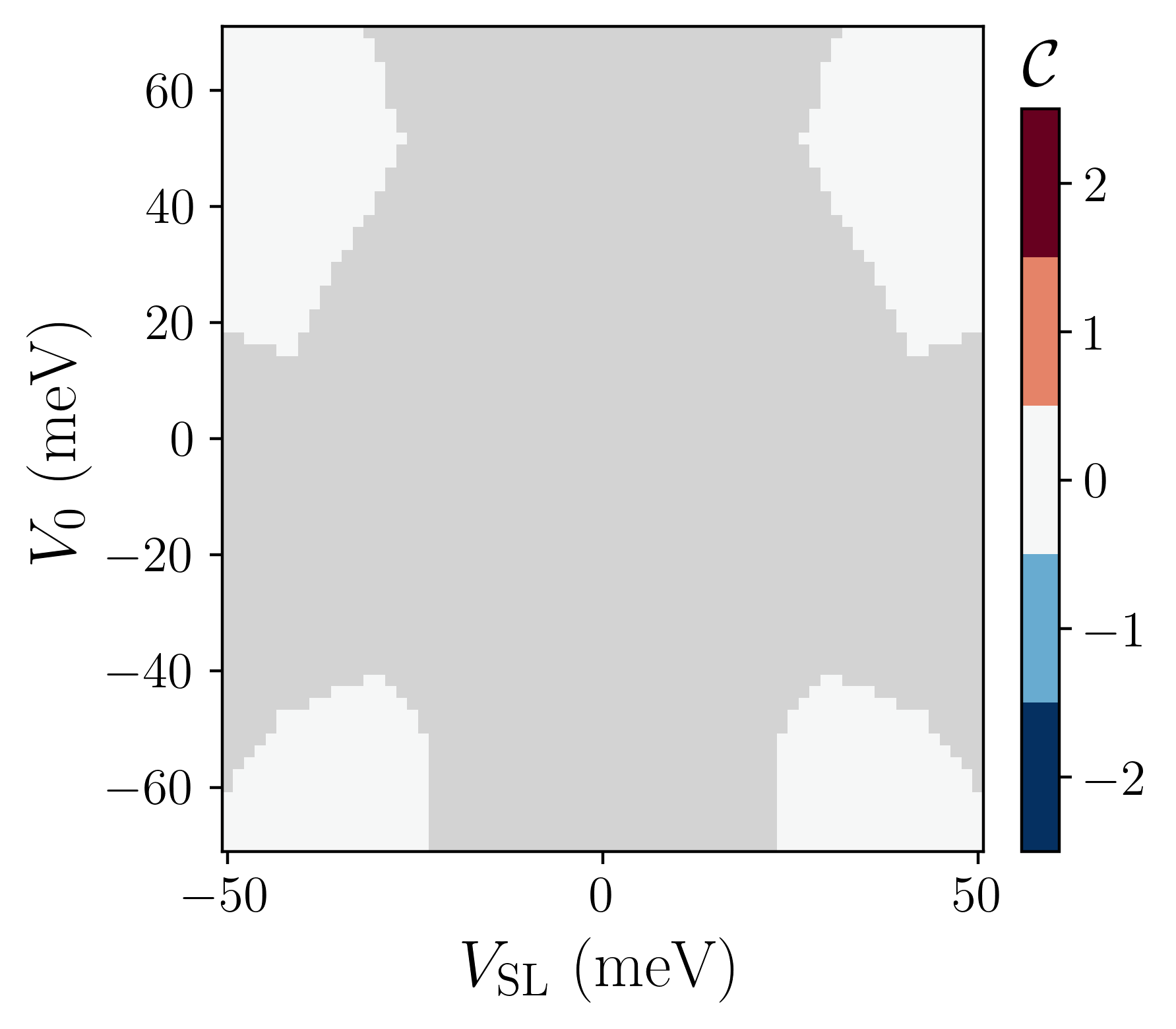}}
    \subfigure[]{\includegraphics[height=0.42\linewidth]{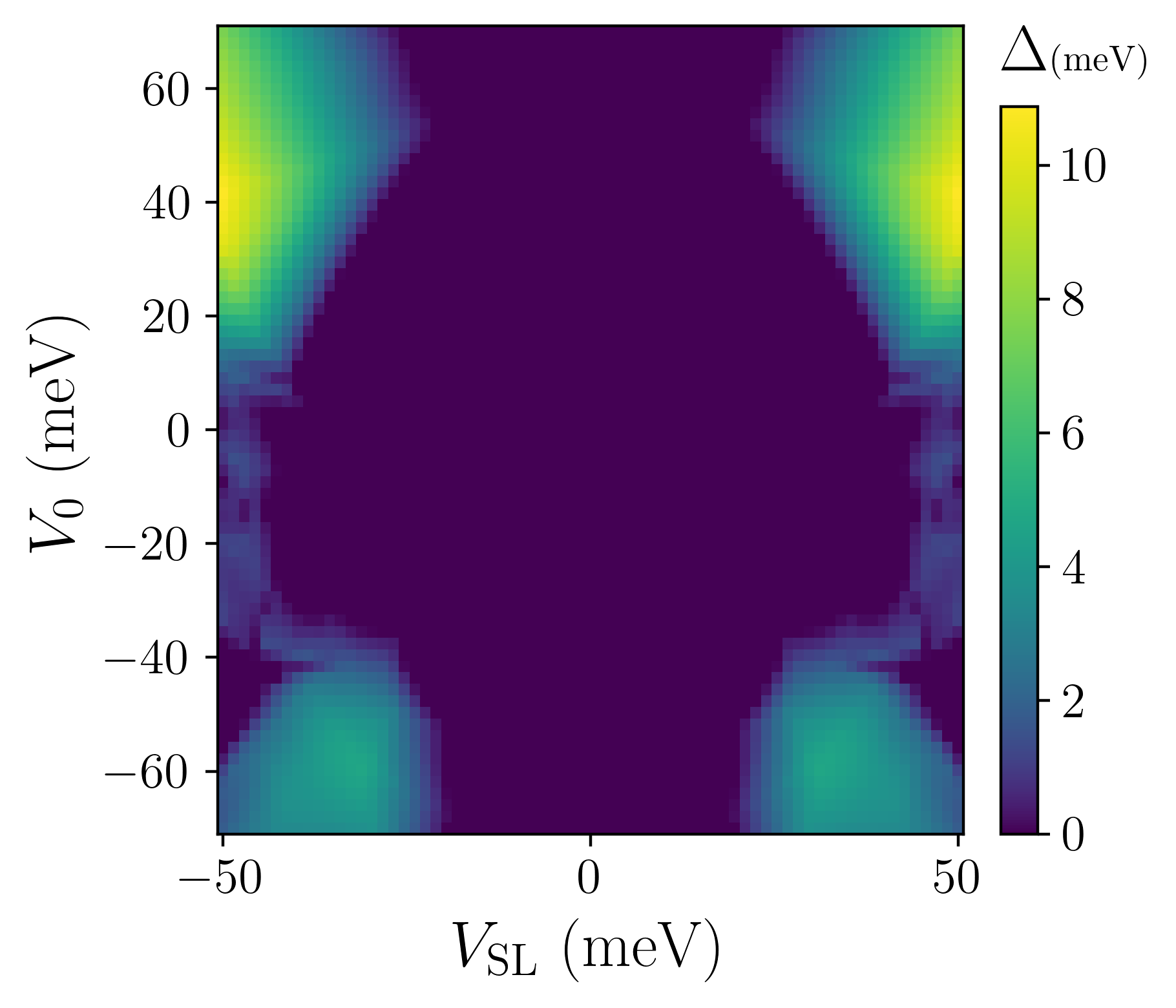}}
    \caption{(a) Chern number $\mathcal{C}$ and (b) band gap $\Delta$ of the $n=1$ band of $L = \SI{30}{nm}$ square sBLG. Grayed regions of (a) indicated that the Chern number is not numerically well-resolved due to the proximity of a band closing.}
    \label{fig:sq-n1-phase}
\end{figure}

We also consider sBLG with a square superlattice geometry. 
The reciprocal lattice vectors are listed below Eq.~(\ref{eq:VSL}).
The band structure for this system is shown in Fig.~\ref{fig:sq-bands}(a) at example parameter values $(V_{\text{SL}}, V_0) = (\SI{30}{meV}, \SI{30}{meV})$. The $n = 0$ band, which in this case is not topological, is highlighted in red, and its Berry curvature distribution is shown in Fig.~\ref{fig:sq-bands}(b). Note that the four-fold rotational symmetry of the superlattice is broken by the trigonal warping of the underlying bilayer graphene dispersion. For $v_3 = 0$, the symmetry is restored.

$\mathcal{C}$ and $\Delta$ as functions of $V_{\text{SL}}$ and $V_0$ are shown for the $n = 0$ and $n = 1$ bands in Fig.~\ref{fig:sq-n0-phase} and Fig.~\ref{fig:sq-n1-phase}, respectively. Note the symmetry of this phase diagram under a reflection $V_{\text{SL}} \to -V_{\text{SL}}$. This is a consequence of the fact that, for a square superlattice, translation in real space by $(L/2, L/2)$ corresponds to an inversion of the superlattice potential, but leaves both $H_{\text{BLG}}$ and the uniform displacement potential $V_0$ unchanged. 

Up to this symmetry, the $n = 0$ band has two topological phases in the scanned region of parameter space, with $\mathcal{C} = +1$ and $\mathcal{C} = +2$. However, both require fine-tuning of $(V_{\text{SL}}, V_0)$ to access this small parameter regime and do not have band gaps exceeding 3.5 meV. 
Thus, we do not consider these regimes to be fruitful regions of phase space in which to search for an FCI. The $n = 1$ band, meanwhile, has no topological phases with $\Delta > \SI{1.5}{meV}$ at all in the scanned region, as shown in Fig.~\ref{fig:sq-n1-phase}.

Based on these results, we conclude that square superlattice geometries are less favorable than triangular geometries for realizing topological flat bands in sBLG.

\section{Dependence on superlattice periodicity}
\label{sec:scale}

\begin{figure}
    \centering
    \subfigure[]{\includegraphics[height=0.42\linewidth]{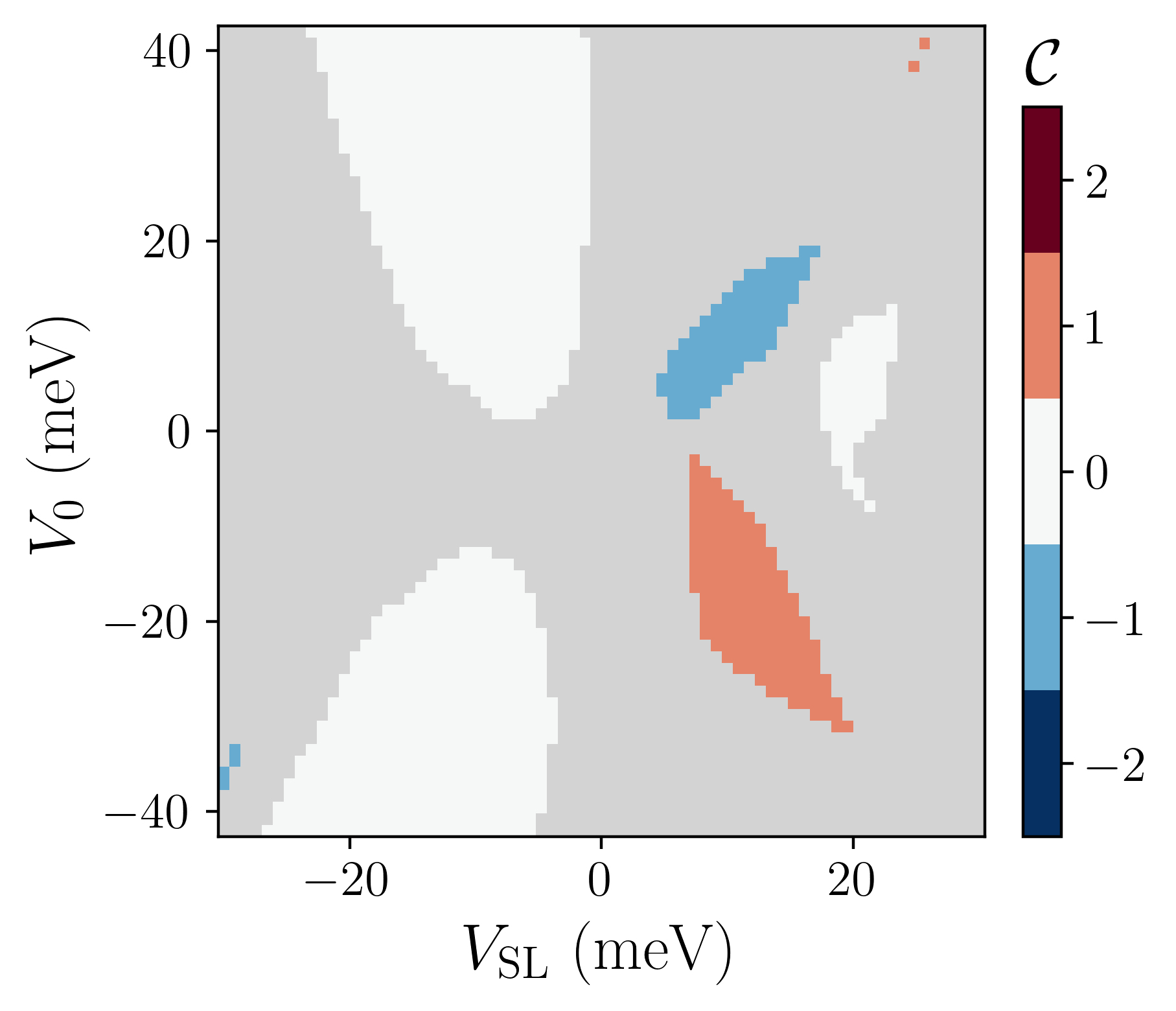}}
    \subfigure[]{\includegraphics[height=0.42\linewidth]{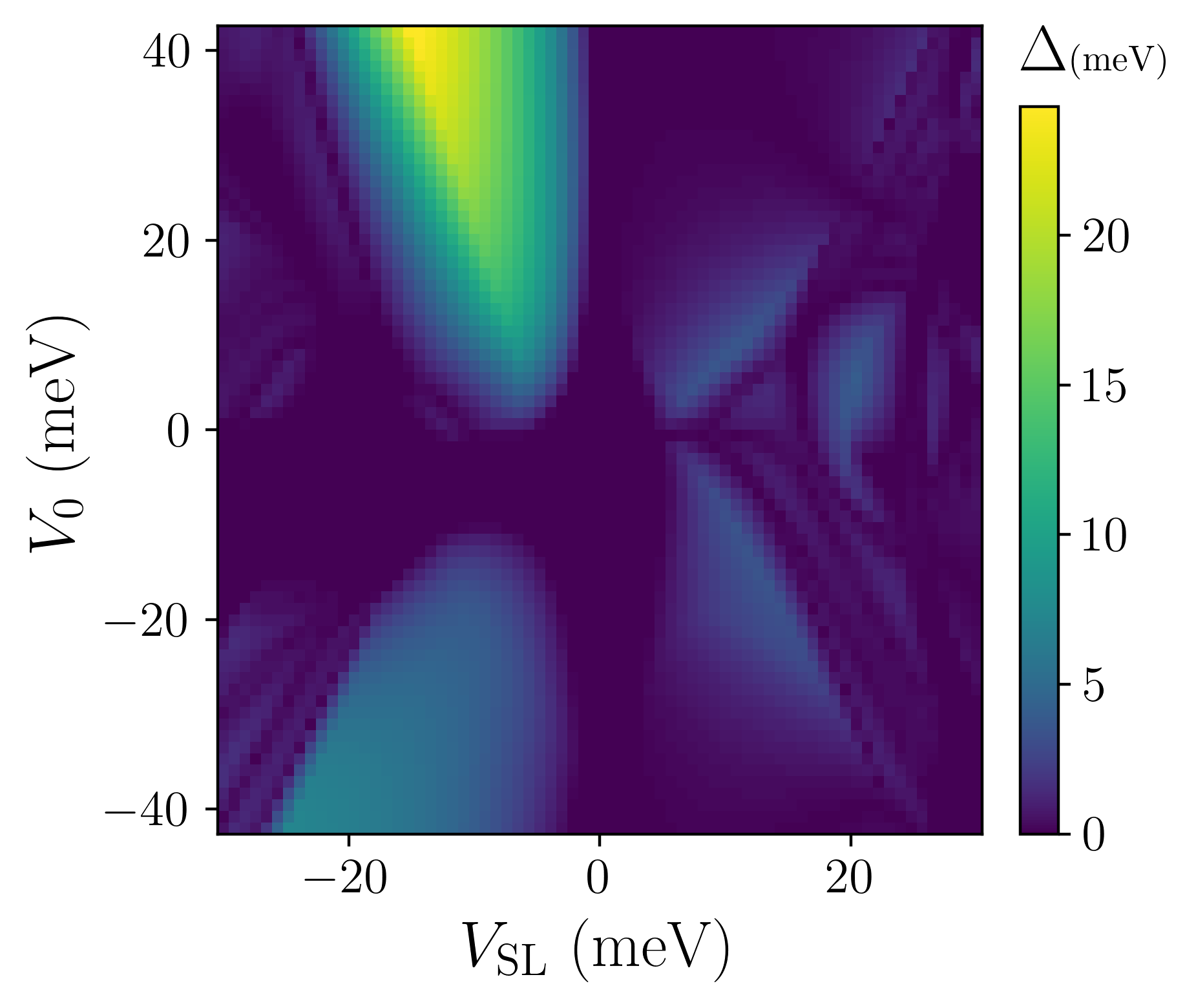}}
    \caption{(a) Chern number $\mathcal{C}$ and (b) band gap $\Delta$ of the $n=0$ band of $L = \SI{50}{nm}$ triangular sBLG. Grayed regions of (a) indicated that the Chern number is not numerically well-resolved due to the proximity of a band closing.}
    \label{fig:triag-50nm-phase}
\end{figure}

\begin{figure}
    \centering
    \subfigure[]{\includegraphics[width=0.49\linewidth]{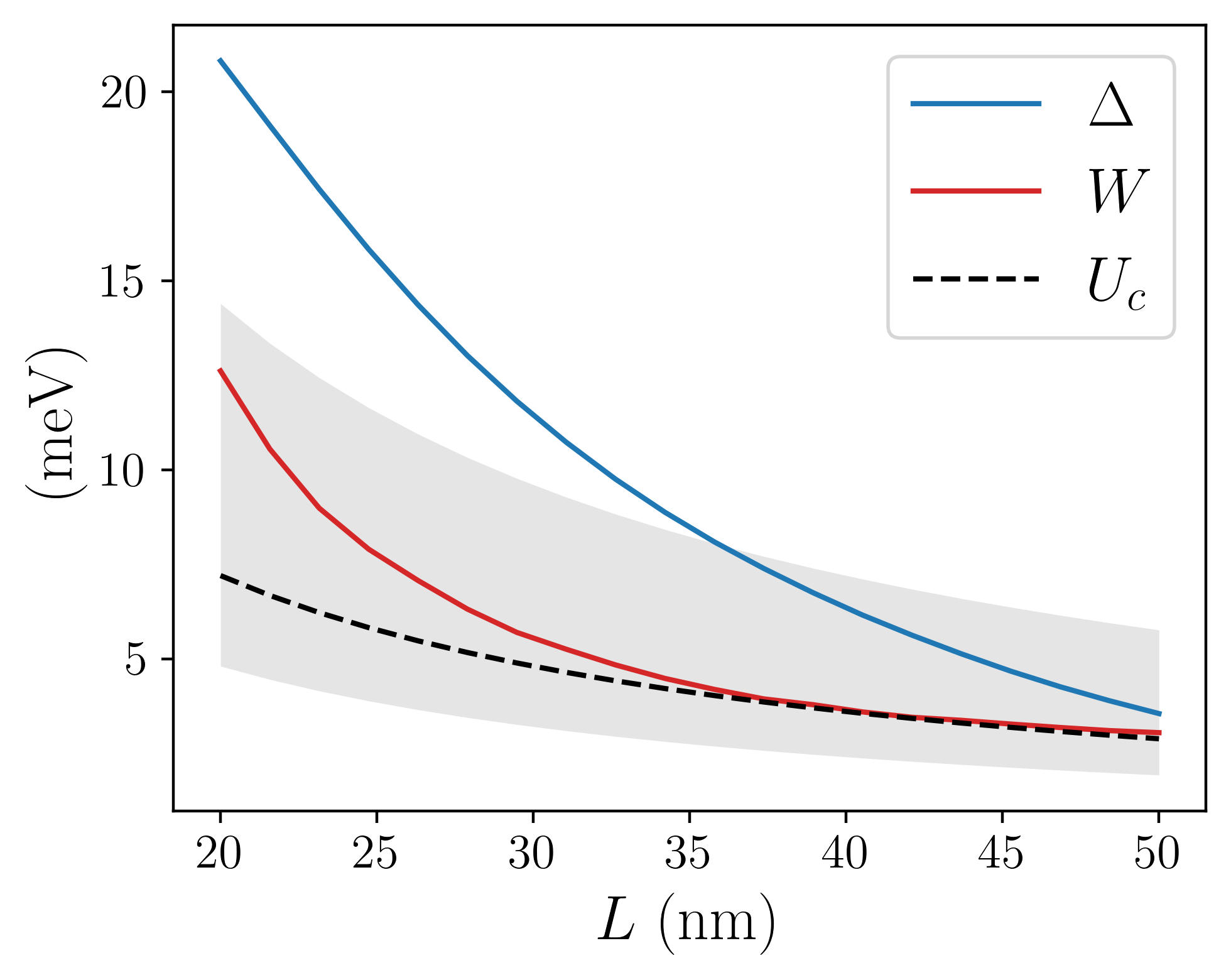}}
    \subfigure[]{\includegraphics[width=0.49\linewidth]{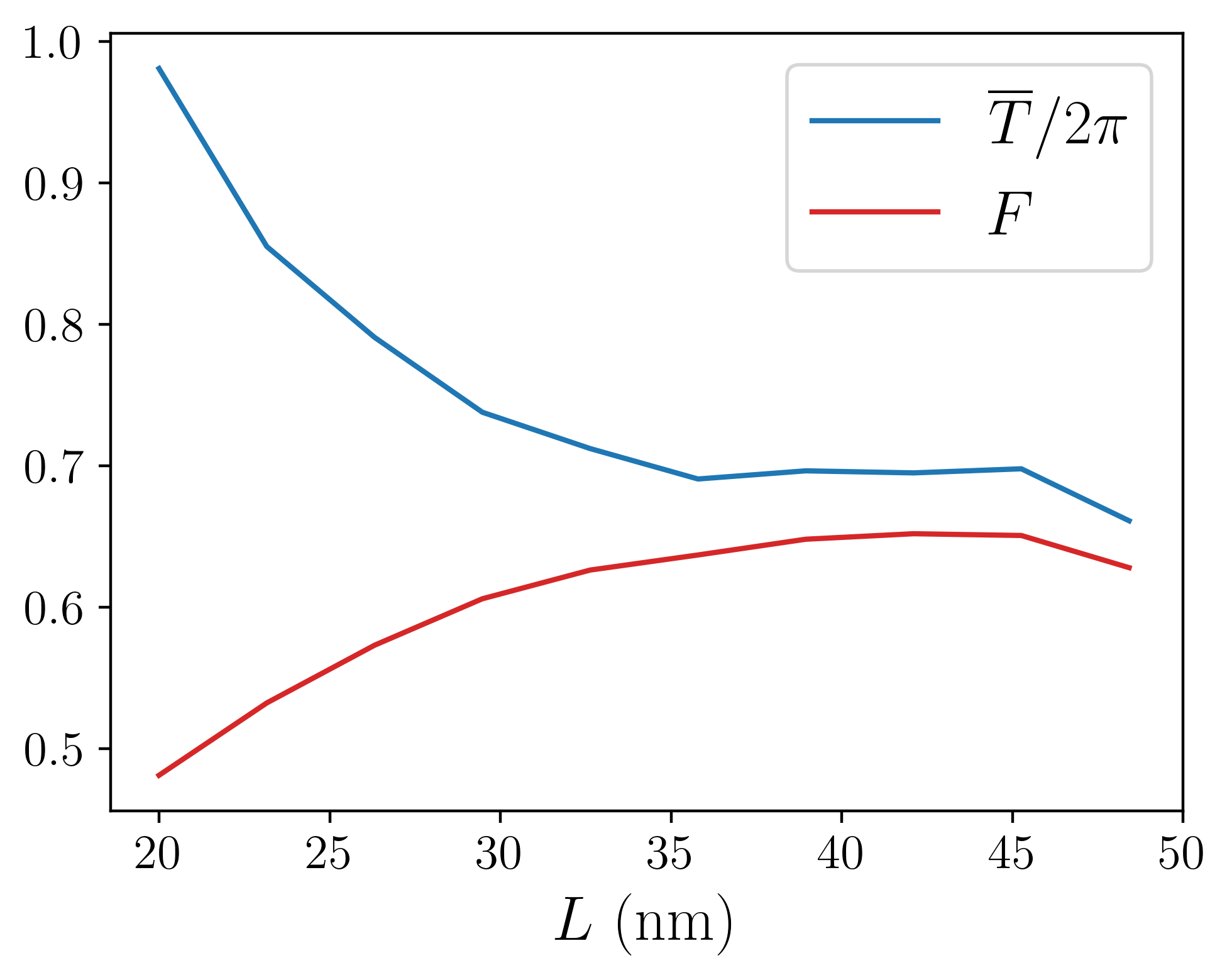}}
    \caption{Indicators as a function of $L$ in triangular sBLG, with $V_{\text{SL}}$ and $V_0$ chosen to maximize $\Delta$ within the $\mathcal{C} = +1$ phase of interest highlighted in Fig.~\ref{fig:triag-n0-phase}. (a) $\Delta$ and $W$, together with an estimate of the Coulomb interaction scale $U_c \sim e^2/4\pi\epsilon L$. The dashed black line corresponds to $\epsilon/\epsilon_0 = 10$ while the shaded gray region corresponds to values between 5 and 15. (b) Quantum geometry indicators $\overline{T}$ (divided by $2\pi$ to fit on a shared axis) and $F$.}
    \label{fig:L-scaling}
\end{figure}

We now consider the effect of rescaling the superlattice constant $L \to aL$. This rescales the Brillouin zone by $\vb{k} \to \vb{k}/a$, and thus Eq.~(\ref{eq:Hblg}) shows that simultaneously rescaling $t \to t/a$ rescales the entire effective Hamiltonian for BLG as $H_{\text{BLG}} \to H_{\text{BLG}}/a$. In other words, changing $L$ is equivalent to an overall change of the energy scale of the system together with a renormalization of the interlayer hopping $t$, as one would predict from dimensional analysis. This means that we generically expect larger values of $L$ to lead to smaller band widths and gaps, and require lesser gate voltages to reach the regions of interest in the phase diagram. The renormalization of $t$ does not substantially change this qualitative picture (compare Fig.~\ref{fig:triag-50nm-phase}, showing $\mathcal{C}$ and $\Delta$ for the $n=0$ band of $L = \SI{50}{nm}$ triangular sBLG, and Fig.~\ref{fig:triag-n0-phase}, which shows the same for $L = \SI{30}{nm}$), but it does have quantitative effects.

The band gap and bandwidth $\Delta$ and $W$ are shown as functions of $L$ in the prominent $\mathcal{C} = +1$ phase of triangular sBLG in Fig.~\ref{fig:L-scaling}(a). 
Since, as we showed in the previous section, $\Delta$ is the most sensitive indicator in this region and the other indicators tend to be reasonably well optimized near its maximum (see Fig.~\ref{fig:triag-n0-zoom}), the parameters $V_{\text{SL}}$ and $V_0$ are selected at each value of $L$ to maximize $\Delta$.
Also shown in Fig.~\ref{fig:L-scaling}(a) is an estimate of the Coulomb interaction scale $U_c \sim e^2/4\pi\epsilon L$. The dashed black line corresponds to $\epsilon/\epsilon_0 = 10$, the same value used in Fig.~\ref{fig:ind-table}, while the shaded gray region corresponds to values between 5 and 15.
Both $\Delta$ and $W$ decrease with $L$ faster than the $1/L$ dependence predicted by naive dimensional analysis, and thus also faster than $U_c$, due to the renormalization of $t$. On the other hand, the relative band flatness $\Delta / W$ is maximized near $L = \SI{30}{nm}$ and quickly decreases for both larger and smaller $L$. 

The quantum geometry indicators $\overline{T}$ and $F$ are shown in Fig.~\ref{fig:L-scaling}(b) as functions of $L$, using the same prescription for $V_{\text{SL}}$ and $V_0$ as above. 
Neither of the quantum geometry markers vary significantly for $L$ between 30 and 50 nm, but as $L$ decreases below 30 nm, $\overline{T}$ trends upward while $F$ trends downward.

\section{Conclusion}
\label{sec:disc}

We calculated the band structure and momentum-space quantum geometry of multiple bands near the Fermi level in both triangular and square sBLG across a wide range of parameter values. This allowed us to evaluate single-particle indicators for FCI stability and identify the most promising regions of the phase diagram for the experimental realization of an FCI. Our main result is that the most prominent topological phases appear in the $n = 0$ band of triangular sBLG, and that these phases exhibit an optimal line in parameter space along which the band gap is sharply maximized and other indicators remain favorable. At a superlattice length scale of $L = \SI{30}{nm}$, this line extends from approximately $(V_{\text{SL}}, V_0) = (\SI{20}{meV}, -\SI{12}{meV})$ to $(\SI{35}{meV}, -\SI{35}{meV})$, with the most optimal values being near $(\SI{30}{meV}, -\SI{30}{meV})$, at which the single-particle indicators for FCI stability are similar to or better than those for magic-angle TBG. Smaller values of $L$ lead to greater band dispersion relative to the strength of the Coulomb interaction as well as less ideal quantum geometry, though band gaps increase and Berry curvature fluctuations decrease. Larger values lead to flatter bands and allow the phases of interest to be accessed with smaller applied gate voltages, but also result in substantially smaller band gaps. 

Our results guide the experimental search for an FCI in sBLG. 
This guide can be expanded in future work by incorporating the effects of higher harmonics of the superlattice potential and considering non-Bravais superlattice geometries such as the kagome lattice.
In addition, our analysis can be extended to other multilayer graphene stacks that realize topological flat bands \cite{ghorashi2023multilayer}.
    
\acknowledgements

The authors thank Xu Du and Daniel Parker for useful conversations and correspondence.
D.A., M.L., and A.D. acknowledge support from the National Science Foundation under the Columbia MRSEC on Precision-Assembled Quantum Materials (PAQM), Grant No. DMR-2011738.
A.D. acknowledges support from the Keele and Jane and Aatos Erkko foundations as part of the SuperC collaboration.
S.A.A.G. acknowledges support from the Air Force Office of Scientific Research under Grant No. FA9550-20-1-0260.
In addition, J.C. acknowledges support from the Alfred P. Sloan Foundation through a Sloan Research Fellowship and from the Flatiron Institute, a division of the Simons Foundation.

\bibliography{bibliography}{}

\end{document}